\documentclass[fleqn,usenatbib]{mnras}

\usepackage{newtxtext,newtxmath}

\usepackage[T1]{fontenc}
\usepackage{ae,aecompl}

\newcommand{\ngts}{NGTS}
\newcommand{\kepler}{{\it Kepler}}
\newcommand{\ktwo}{{\it K2}}

\newcommand{\tess}{{\it TESS}}

\newcommand{\gaia}{{\it Gaia}}

\newcommand{\Mjup}{\mbox{M$_{\rm J}$}}
\newcommand{\msun}{\mbox{M$_{\odot}$}}

\newcommand{\feh}{[Fe/H]}
\newcommand{\ebv}{E(B-V)}
\newcommand{\vsini}{$v \sin i$}

\newcommand{\blanco}{Blanco 1}
\newcommand{\pleiades}{Pleiades}

\newcommand{\gacf}{G-ACF}

\newcommand{\colblue}[1]{\textcolor{blue}{#1}}

\newcommand{\gmk}{G\,--\,K$_{\rm s}$}
\newcommand{\igmk}{(G\,--\,K$_{\rm s}$)$_{0}$}
\newcommand{\bprp}{G$_{\rm BP}$\,--\,G$_{\rm RP}$}
\newcommand{\ibprp}{(G$_{\rm BP}$\,--\,G$_{\rm RP}$)$_{0}$}
\newcommand{\bp}{G$_{\rm BP}$}
\newcommand{\rp}{G$_{\rm RP}$}
\newcommand{\mg}{M$_{\rm G}$}
\newcommand{\mk}{M$_{\rm K_{s}}$}
\newcommand{\ks}{K$_{\rm s}$}

\newcommand{\nstars}{170}
\newcommand{\nProt}{127}
\newcommand{\nProtsame}{118}
\newcommand{\nProtdiff}{9}
\newcommand{\nphotbin}{39}

\usepackage{graphicx}	
\usepackage{amsmath}	
\usepackage{amssymb}	
\usepackage{xcolor}
\usepackage{pdflscape}

\title[NGTS clusters survey. I. Rotation in Blanco 1]{NGTS clusters survey. I. Rotation in the young benchmark open cluster Blanco 1}

\author[E. Gillen et al.]{
\parbox{\textwidth}{
Edward Gillen,$^{1,\dagger}$\thanks{E-mail: ecg41@cam.ac.uk (EG)}
Joshua T. Briegal,$^{1}$
Simon~T.~Hodgkin,$^{2}$ 
Daniel Foreman-Mackey,$^{3}$
Floor Van Leeuwen,$^{2}$
James A. G. Jackman,$^{4,5}$
James McCormac,$^{4,5}$
Richard G. West,$^{4,5}$
Didier Queloz,$^{1}$
Daniel Bayliss,$^{4,5}$
Michael R. Goad,$^{6}$
Christopher A. Watson,$^{7}$ 
Peter J.\ Wheatley,$^{4,5}$
Claudia Belardi,$^{6}$
Matthew R. Burleigh,$^{6}$
Sarah L. Casewell,$^{6}$
James S. Jenkins,$^{8,9}$
Liam Raynard,$^{6}$
Alexis M. S. Smith,$^{10}$
Rosanna H. Tilbrook$^{6}$
and Jose I. Vines$^{8}$
}
\vspace{4mm}\\
$^{1}$Astrophysics Group, Cavendish Laboratory, J.J. Thomson Avenue, Cambridge CB3 0HE, UK\\ 
$^{2}$Institute of Astronomy, University of Cambridge, Madingley Rise, Cambridge CB3 0HA, UK\\
$^{3}$Center for Computational Astrophysics, Flatiron Institute, New York, NY\\
$^{4}$Dept.\ of Physics, University of Warwick, Gibbet Hill Road, Coventry CV4 7AL, UK\\
$^{5}$Centre for Exoplanets and Habitability, University of Warwick, Gibbet Hill Road, Coventry CV4 7AL, UK\\
$^{6}$Department of Physics and Astronomy, University of Leicester, Leicester, LE1 7RH, UK\\
$^{7}$Astrophysics Research Centre, School of Mathematics and Physics, Queen's University Belfast, Belfast, BT7 1NN, UK\\
$^{8}$Departamento de Astronomia, Universidad de Chile, Casilla 36-D, Santiago, Chile\\
$^{9}$Centro de Astrof\'isica y Tecnolog\'ias Afines (CATA), Casilla 36-D, Santiago, Chile.\\
$^{10}$Institute of Planetary Research, German Aerospace Center, Rutherfordstr. 2, 12489 Berlin, Germany\\
$^{\dagger}$Winton Fellow
}

\date{Accepted 2019 October 03. Received 2019 October 03; in original form 2019 March 07}

\pubyear{2019}

\begin{document}
\label{firstpage}
\pagerange{\pageref{firstpage}--\pageref{lastpage}}
\maketitle

\begin{abstract}

We determine rotation periods for \nProt\ stars in the $\sim$115 Myr old \blanco\ open cluster using $\sim$200 days of photometric monitoring with the Next Generation Transit Survey (\ngts). These stars span F5--M3 spectral types (1.2\,$\gtrsim$\,$M$\,$\gtrsim$\,0.3\,\msun) and increase the number of known rotation periods in \blanco\ by a factor of four. We determine rotation periods using three methods: Gaussian process (GP) regression, generalised autocorrelation (\gacf) and Lomb-Scargle (LS) periodograms, and find that GPs and \gacf\ are more applicable to evolving spot modulation patterns. 
Between mid-F and mid-K spectral types, single stars follow a well-defined rotation sequence from $\sim$2 to 10 days, whereas stars in photometric multiple systems typically rotate faster. 
This may suggest that the presence of a moderate-to-high mass ratio companion inhibits angular momentum loss mechanisms during the early pre-main sequence, and this signature has not been erased at $\sim$100 Myr. 
The majority of mid-F to mid-K stars display evolving modulation patterns, whereas most M stars show stable modulation signals. 
This morphological change coincides with the shift from a well-defined rotation sequence (mid-F to mid-K stars) to a broad rotation period distribution (late-K and M stars). Finally, we compare our rotation results for \blanco\ to the similarly-aged \pleiades: the single star populations in both clusters possess consistent rotation period distributions, which suggests that the angular momentum evolution of stars follows a well-defined pathway that is, at least for mid-F to mid-K stars, strongly imprinted by $\sim$100 Myr.

\end{abstract}




\section{Introduction}

The initial mass, composition and angular momentum of a star define much of its evolutionary pathway. Rotation influences the internal structure, mixing and energy transport in stars, as well as driving the stellar dynamo, which in turn gives rise to starspots, high energy radiation and stellar winds \citep{Bouvier14}.

At the start of the pre-main sequence (PMS), solar-type and low-mass stars ($M > 0.3$\,\msun) typically have rotation periods between 1--10 days and show a bimodal rotation distribution with peaks at $\sim$2 and 8 days \citep{Herbst01}. This bimodality is usually attributed to the presence of circumstellar disks, with the slow rotators thought to be prevented from spinning up due to ongoing interaction with their disks, whereas the fast rotators are believed to have already dissipated their inner disks and hence are spinning up towards the zero age main sequence (ZAMS) \citep{Barnes03,Bouvier13}. 
Rotation rates increase towards the ZAMS, where stars of a given mass arrive with a range of rotation velocities \citep[e.g.][]{Stauffer87a,Terndrup00,Rebull16,Rebull16a}.

During the main sequence (MS) stage of evolution, rotation rates slow and stars of a given mass converge to rotate with a characteristic period that increases with time. 
The time for this convergence is mass dependent: for stars with convective envelopes, higher mass stars converge faster than their lower mass counterparts \citep{Stauffer87}. During this evolution, angular momentum is lost through magnetised stellar winds \citep[e.g.][]{Chaboyer95a,Chaboyer95b,Reiners12a} and redistributed within the stellar interior \citep[e.g.][]{Eggenberger05,Lagarde12,Charbonnel13}, which leaves older stars with weaker magnetic fields and lower levels of high energy radiation \citep[e.g.][and references therein]{Vidotto14,Johnstone15}. Recent theoretical models now reproduce the main PMS and MS evolutionary trends in the observed rotation behaviour of solar-type and low-mass stars \citep{Gallet13,Gallet15}.

Young open clusters offer a particularly useful tool for understanding the evolution of stellar rotation during the first billion years ($t<1$\,Gyr). Open clusters are populations of stars that span a range of masses but possess essentially the same age and composition. Observational studies of rotation in young open clusters date back to the 1960s, but rapidly expanded as large-scale photometric surveys arose using wide-field cameras on small-to-medium class telescopes, e.g.: the Monitor Survey \citep{Irwin06}, HATNet \citep{Hartman10}, SuperWASP \citep{Delorme11}, KELT \citep{Cargile14}, PTF \citep{Covey05} and \ktwo\ \citep{Rebull16,Rebull16a,Rebull17,Stauffer16,Douglas17,Douglas19}\footnote{HATNet = Hungarian-made Automated Telescope Network; SuperWASP = Super Wide Angle Search for Planets; KELT = The Kilodegree Extremely Little Telescope; PTF = Palomar Transient Factory; \ktwo\ = \kepler/\ktwo.}.

Rotation periods can be determined from photometric monitoring of young active stars by tracking the brightness modulation patterns that arise from the longitudinal inhomogeneity of surface starspot distributions as the stars rotate. In principle, this is relatively straightforward, but complications arise from the fact that starspots appear, evolve in both size and position, and disappear over many rotation periods. The complexity of methods used to estimate rotation periods from stellar light curves, therefore, should be matched to the precision and duration of the data analysed.

Combining rotation period information from clusters of different ages allows us to probe the evolution of stellar rotation as a function of age. Pairs of similarly-aged clusters are particularly valuable, as they offer a means to determine the rotation period distribution at a given age from two independent samples of stars in different cluster environments. Such pairs are rare, especially at older ($t>100$\,Myr) ages, with the two main examples being the Hyades and Praesepe (both 700--800 Myr; \citealt{Brandt15,Brandt15a}); and the Pleiades and \blanco\ (both 100--120 Myr; \citealt{Babusiaux18}, hereafter \colblue{B18})\footnote{We note that M35 and NGC\,2516 have traditionally both been assigned ages of $\sim$150\,Myr \citep{Meibom09,Irwin07a}, but the Gaia DR2 estimated age of NGC\,2516 from its HR diagram in \colblue{B18} is $\sim$300\,Myr. Without a comparable DR2-derived age for M35 in \colblue{B18} we refrain from labelling them as similarly-aged here.}. The Pleiades, Hyades and Praesepe all have a long history of rotation studies, culminating in recent observations by \kepler/\ktwo\ spanning 75 days. \blanco, however, lacks such a precise long-term photometric monitoring campaign. This motivated the Next Generation Transit Survey (NGTS; \citealt{Chazelas12,Wheatley18}) to observe \blanco\ with precise mmag photometry spanning $\sim$200 days; these observations form the basis of this paper.

\blanco\ is a $\sim$115\,Myr old Galactic open cluster, situated in the local spiral arm at a distance of $\sim$240\,pc in the direction towards and below the Galactic centre (\colblue{B18}). It is comprised of 489 Gaia DR2-confirmed stars, ranging from B-to-M spectral types, as well as tens of likely brown dwarf members down to $\sim$30\,\Mjup\ \citep[\colblue{B18};][]{Moraux07,Casewell12}. The cluster has a solar metallicity \citep[\feh\,$\sim$\,0.03;][]{Netopil16}, an on-sky stellar density of $\sim$30 stars\,pc$^{-2}$ \citep{Moraux07} and a low reddening along the line of sight (\ebv\,$\sim$\,0.010; \colblue{B18}). Given these properties, \blanco\ is much like a scaled-down version of the Pleiades ($\sim$110\,Myr, 1326 Gaia DR2 members, on-sky stellar density $\sim$65 stars\,pc$^{-2}$, \feh\,$\sim$\,-0.01; \citealt{Moraux03}, \colblue{B18}). Its main outstanding property is its high Galactic latitude ($b=-79\degr$), especially given its moderate proper motion and UVW velocities, which hint at an unusual formation/evolution history compared to most young open clusters (normally located close to the Galactic plane).

\blanco\ was first noted by \citet{Blanco49} and has been extensively studied over the last 70 years. Initial photometric observations identified many of the higher mass members \citep[e.g.][]{Westerlund63,Epstein68,deEpstein85,Westerlund88}, with subsequent photometric and astrometric studies revealing the lower mass population \citep[e.g.][]{Moraux07,Platais11,Casewell12}. 
Spectroscopic studies of \blanco\ have broadly characterised the radial velocity distribution of the cluster \citep[e.g.][]{Mermilliod08,Mermilliod09,Gonzalez09} and provided \vsini\ measurements of the brighter moderate-to-rapid rotators in the cluster. The rotation of stars in \blanco\ has been studied by \citet{Cargile14}, who report photometric rotation periods for 33 stars with spectral types between late-A\,/\,early-F and mid-K. \blanco\ has also been the subject of X-ray surveys \citep{Micela99,Pillitteri03,Pillitteri04,Pillitteri05}, as well searches for debris disks \citep{Stauffer10} and stellar flares \citep{Leitzinger14}. Most recently, Gaia DR2 produced a homogeneous membership list for \blanco\ spanning the cluster's full stellar sequence (\colblue{B18}); we use this Gaia DR2 membership list in the present study. 

This paper presents a study of rotation in \blanco\ using $\sim$200 days of ground-based data from the Next Generation Transit Survey (\ngts). We introduce the NGTS observations in \S \ref{sec:observations}. In \S \ref{sec:get_Prot} we estimate rotation periods using three methods: Gaussian process (GP) regression, generalised autocorrelation (\gacf) and Lomb-Scargle periodograms, and then compare their predictions. In \S \ref{sec:identify_multiples} we identify likely multiple star systems using colour magnitude diagrams (CMDs). We then discuss our rotation periods for \blanco\ in \S \ref{sec:B1_rot}, before comparing to the \pleiades\ in \S \ref{sec:B1_plei_comp}. We conclude in \S \ref{sec:conclusions}.


\section{\ngts\ observations of Blanco 1}
\label{sec:observations}

\begin{figure}
  \centering
  \includegraphics[width=\linewidth]{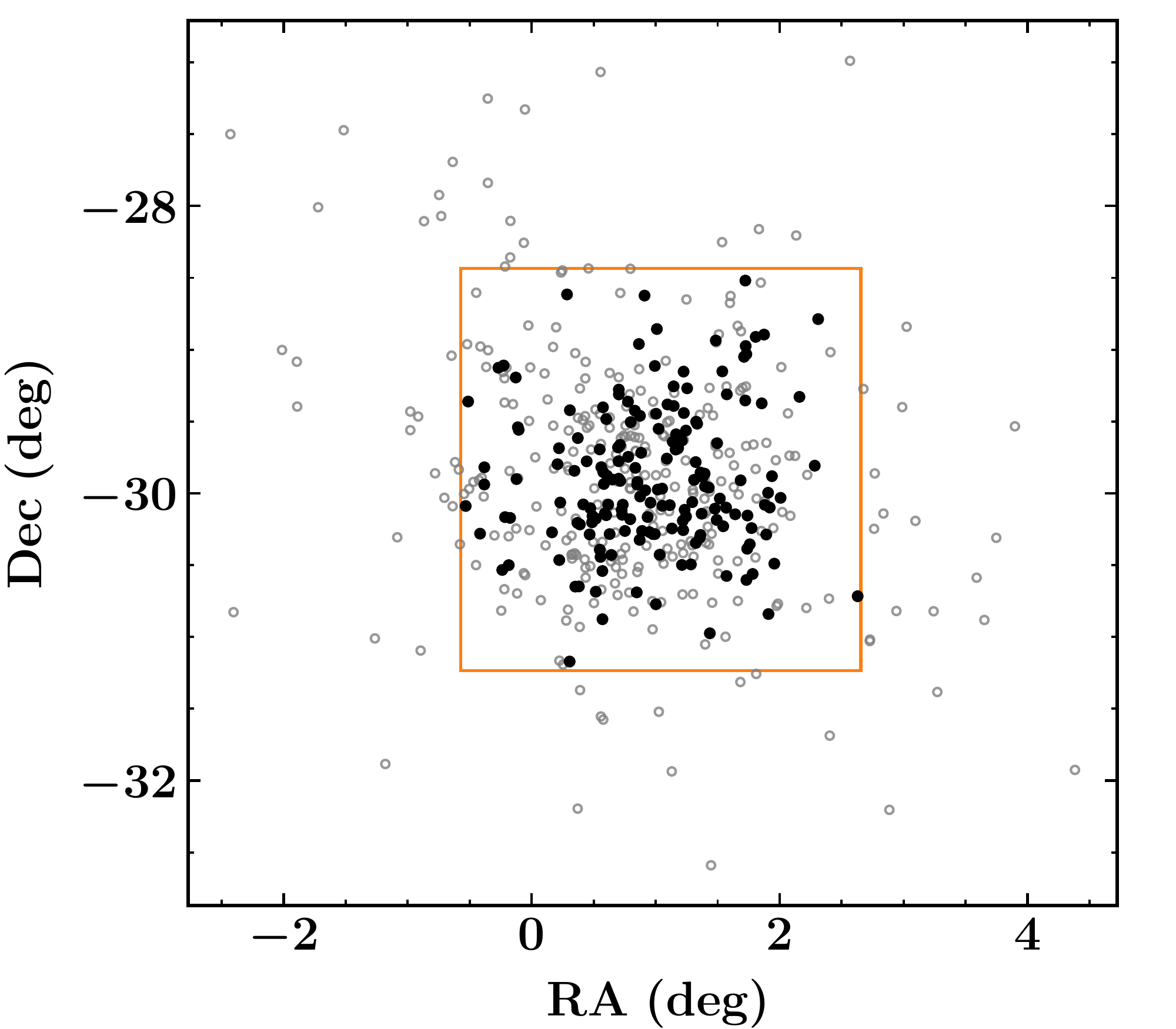}
    \caption{Spatial distribution of \blanco\ members from \citet{Babusiaux18}. The orange box indicates the \ngts\ field of view (FoV), filled black circles represent stars with \ngts\ light curves, and open grey circles indicate stars that were either too faint or fell outside the \ngts\ FoV.}
    \label{fig:spatial}
\end{figure}

\begin{figure}
  \centering
  \includegraphics[width=\linewidth]{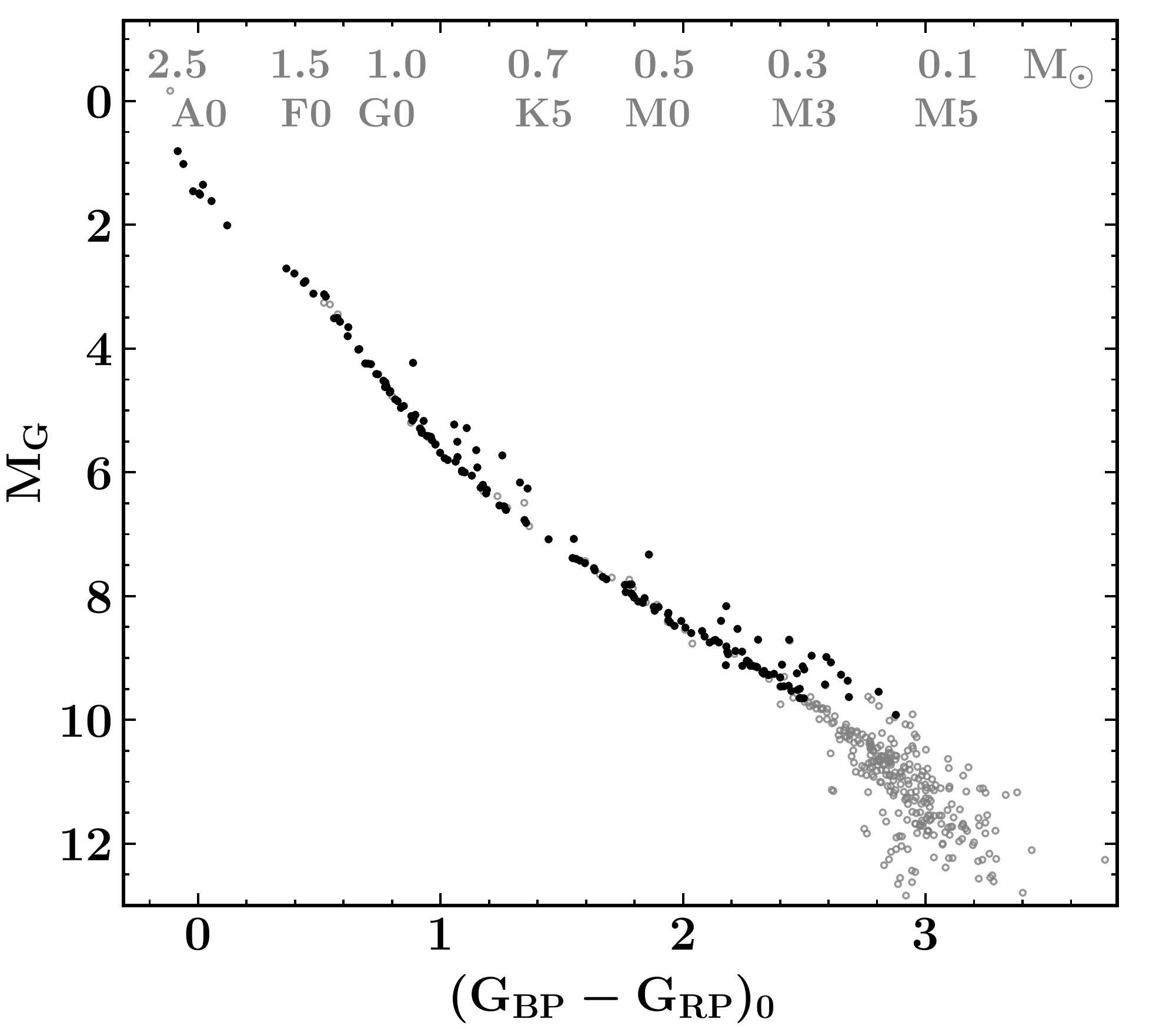}
    \caption{Absolute \gaia\ G vs. \bprp\ colour-magnitude diagram (CMD) for members of \blanco\ from (\citealt{Babusiaux18}; open grey circles) highlighting stars observed by \ngts\ (filled black circles). The \gaia\ photometry and parallaxes reveal a tight cluster sequence with a scattering of likely multiple star systems lying above the single star sequence. For reference, an equal mass binary produces a 0.75 mag excess. The \bprp\ colours have been dereddened assuming E(B-V) = 0.010 for the cluster (\colblue{B18}). \ngts\ observed essentially all cluster members down to a spectral type of $\sim$M3. The stellar masses are MIST model predictions \citep{Dotter16,Choi16} evaluated at the age of \blanco, and the spectral types were estimated using updated information from \citet{Pecaut13} (E. Mamajek online table; see text).}
    \label{fig:CMD}
\end{figure}

The Next Generation Transit Survey (\ngts) comprises twelve 20-cm wide-field roboticised telescopes situated at the ESO Paranal Observatory in Chile. The facility is optimised to detect small exoplanets orbiting K and early M stars \citep[e.g.][]{Bayliss18,West19}, and is designed to achieve milli-mag (mmag) photometric precision across each camera's 2.8$\degr$ field of view (FoV). 

Blanco 1 was observed using a single NGTS camera over a 195-night baseline between 7 May and 18 November 2017. 201\,773 exposures were obtained, at 13-second cadence (with 10\,s exposures), on 134 nights within this period. Of the 489 \blanco\ members from \colblue{B18}, the NGTS FoV encompassed 429 stars (88\% of the cluster members). 170 of these stars had an apparent magnitude brighter than 16 mag in the \ngts\ band, and therefore had photometry automatically extracted by the \ngts\ pipeline, before being binned to 30 minute cadence for this work. The \ngts\ band covers the 520 to 890 nm range, and is therefore similar to a combined R+I filter. We refer the interested reader to \citet{Wheatley18} for further details on the \ngts\ filter and pipeline.

Figures \ref{fig:spatial} and \ref{fig:CMD} show the spatial and colour-magnitude distributions of \blanco\ members, highlighting stars with \ngts\ light curves. Given the mass segregation within the cluster and \ngts's brightness range, nearly all stars in the cluster down to a spectral type of $\sim$M3 have \ngts\ light curves. The spectral types in Figure \ref{fig:CMD} (as well as other Figures and Tables presented here) were estimated using updated information from \citet{Pecaut13}\footnote{\url{http://www.pas.rochester.edu/~emamajek/EEM\_dwarf\_UBVIJHK\_colors\_Teff.txt}} based on their intrinsic \gmk\ and \bprp\ colours, i.e. \igmk\ and \ibprp.


\section{Estimating rotation periods}
\label{sec:get_Prot}

\begin{figure*}
	\includegraphics[width=\columnwidth]{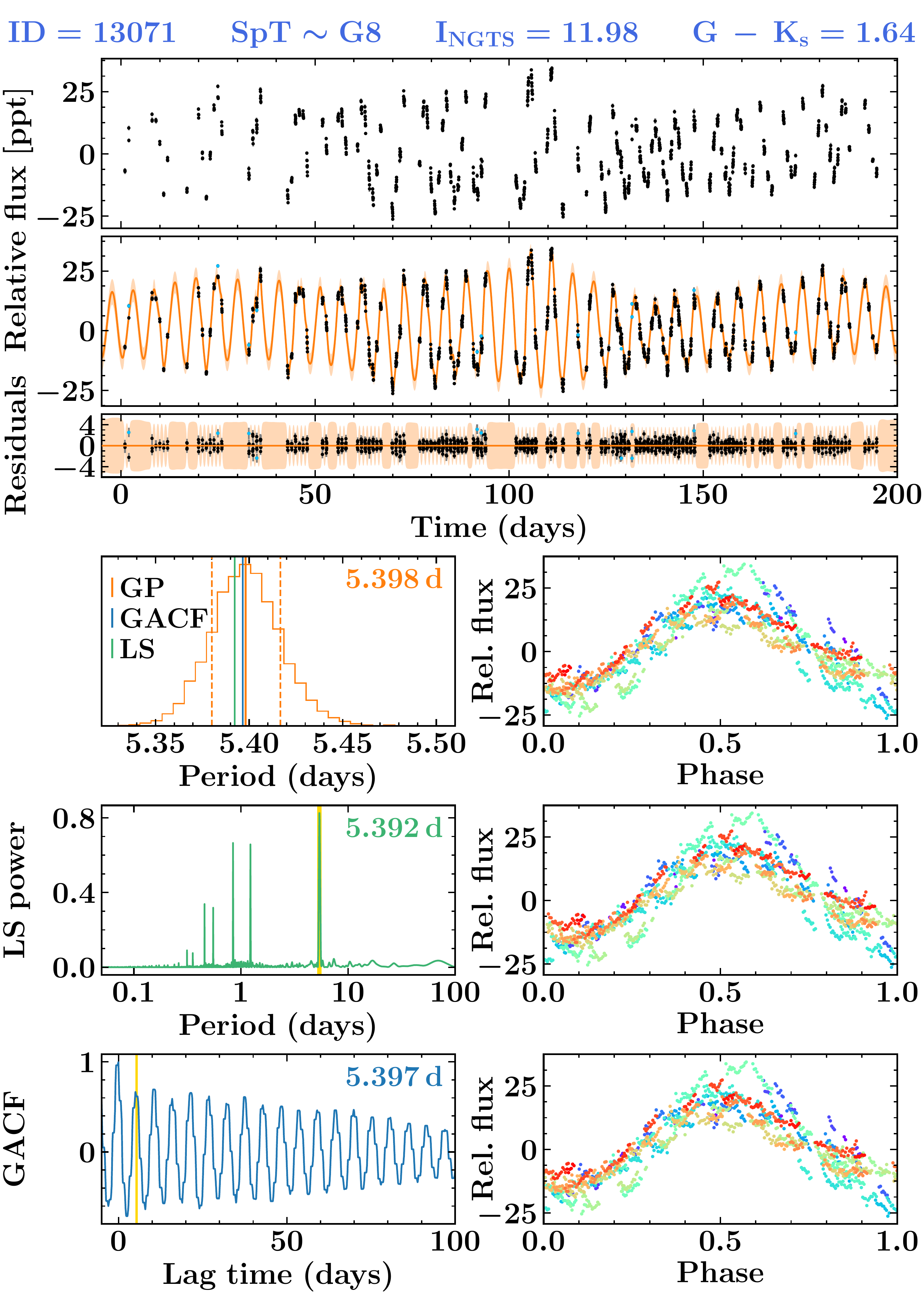}
	\hfill
	\includegraphics[width=\columnwidth]{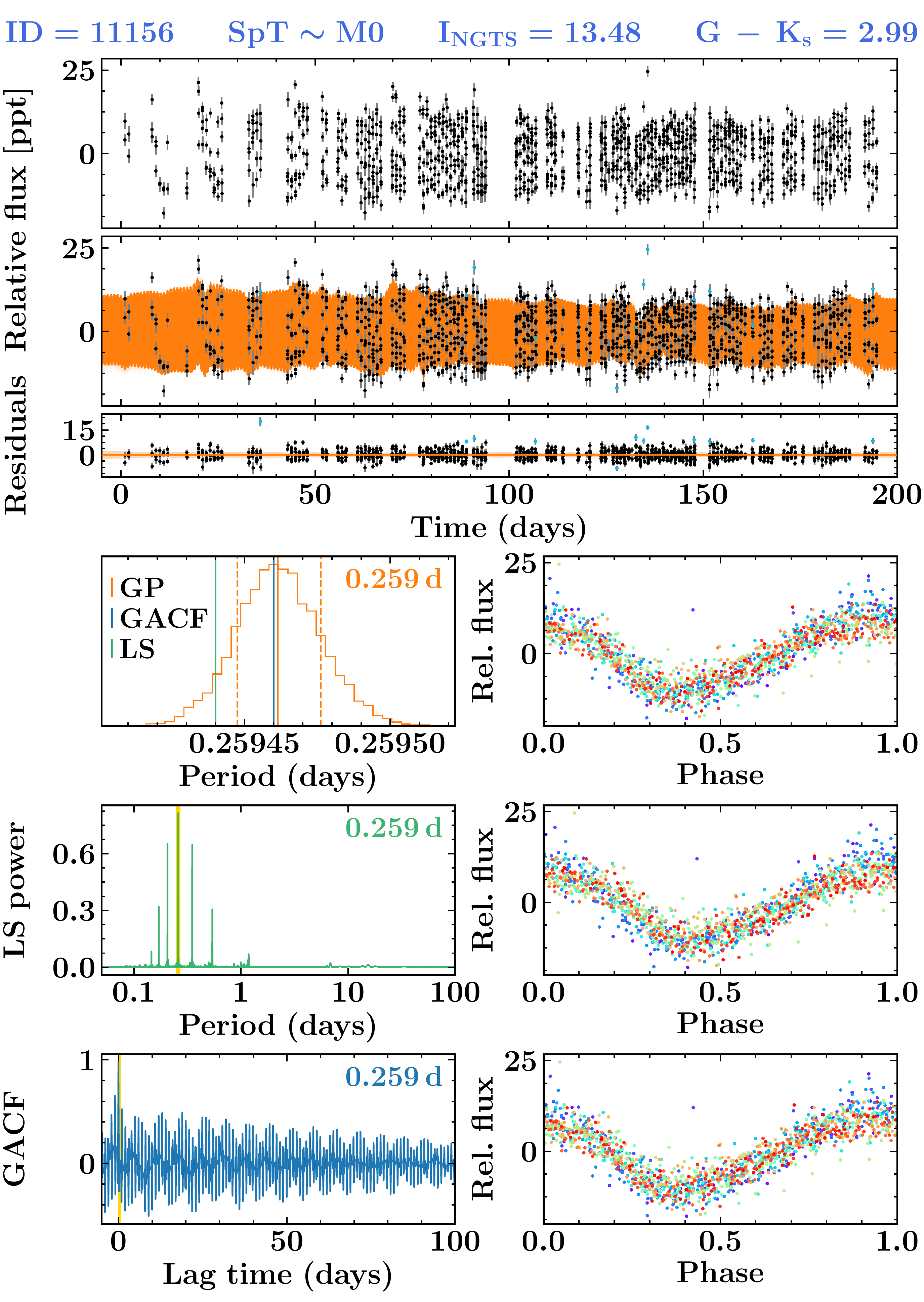}
    \caption{NGTS light curves and period predictions using GPs, LS and \gacf, for two stars (object IDs 13071 and 11156, left and right respectively). 
    In each case, the \emph{top three plots} show the relative flux NGTS light curve in units of parts per thousand (ppt; top), the NGTS light curve with the maximum a posteriori (MAP) GP model (middle) and residuals (bottom). The orange line and shaded region show the mean and 1$\sigma$ uncertainty on the MAP GP model. In the residuals plot, blue points indicate outlying data that was masked by the GP during the fit.
    The \emph{bottom six plots} show the period estimation results (left column) for GPs (top), LS (middle) and \gacf\ (bottom), along with the NGTS data phase-folded on each method's period (right column).
    \emph{Top left}: 1D GP posterior period distribution (orange) with the median period and 1\,$\sigma$ uncertainties (solid and dashed orange lines) shown and the period printed top right. For comparison, the period predictions of \gacf\ and LS are shown by the vertical blue and green solid lines, respectively.
    \emph{Middle left}: LS periodogram in green with the identified period highlighted in yellow and printed top right. 
    \emph{Bottom left}: \gacf\ autocorrelation function in blue (positive direction shown only) with the identified period highlighted in yellow and printed top right. 
    \emph{Right column}: NGTS light curve phase-folded on the corresponding period (GP, LS and \gacf, top-to-bottom), with the rainbow colour scheme indicating data from the beginning (indigo) to the end (red) of the observations.
    The modulation patterns of these two stars are relatively stable during the 200 days of observations and hence the period predictions from GPs, LS and \gacf\ all agree.}
    \label{fig:P_agree}
\end{figure*}

\begin{figure*}
	\includegraphics[width=\columnwidth]{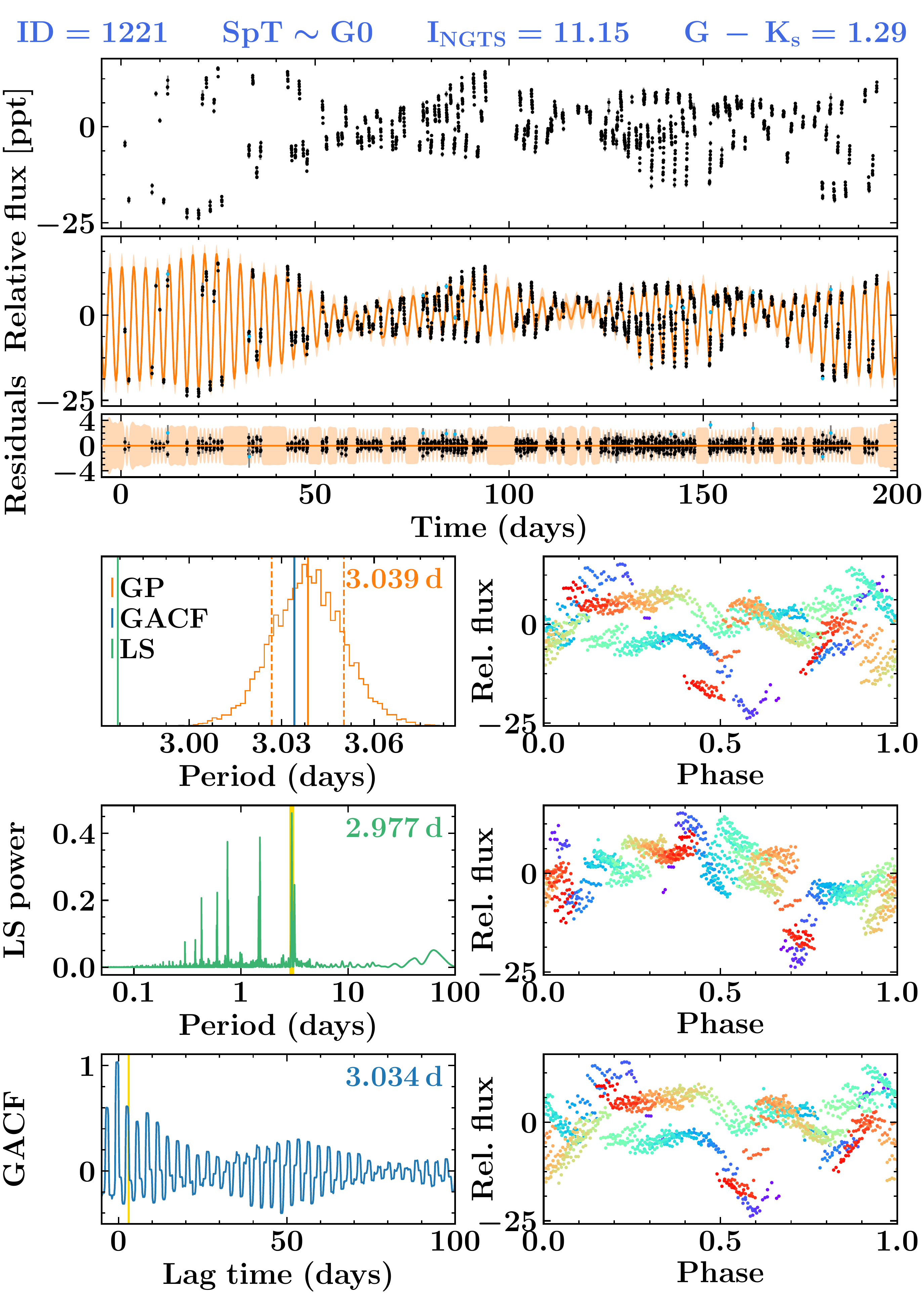}
	\hfill
	\includegraphics[width=\columnwidth]{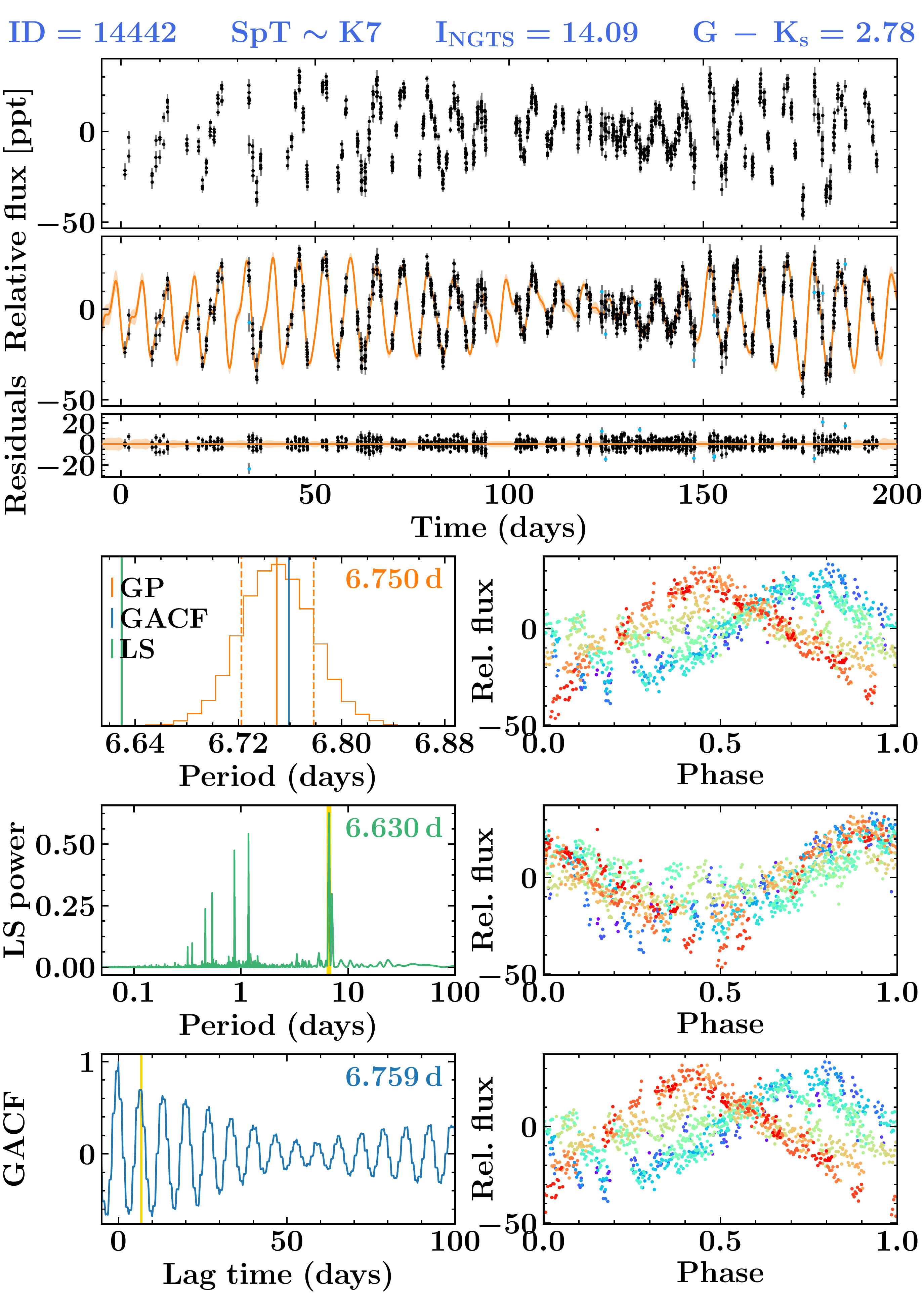}
    \caption{Same as Figure \ref{fig:P_agree} but for objects 1221 and 14442, two stars whose modulation patterns evolve significantly during the 200 day observations. The GP and \gacf\ predictions agree well for these stars, given that they can account for the evolution in the modulation pattern, but the LS prediction is offset because it finds the period that is best fit by a non-evolving sine wave, which is not appropriate for these stars.}
    \label{fig:P_diff}
\end{figure*}

We test three methods: Lomb-Scargle (LS), generalised autocorrelation (G-ACF) and Gaussian process (GP) regression. These differ in their assumptions and complexity, and hence in their appropriateness for estimating rotation periods from photometric rotational modulation. We introduce each of these methods below (\S \ref{sec:LS}, \ref{sec:GACF} and \ref{sec:GPs}) before comparing their assumptions and predictions in \S \ref{sec:P_comp}.

\subsection{Lomb-Scargle}
\label{sec:LS}

The Lomb-Scargle (LS) periodogram is a standard method for detecting periodic signals in unevenly sampled data (see \citealt{VanderPlas17} for a detailed discussion). It has been widely used to estimate stellar rotation periods from stars in the field \citep[e.g.][]{Nielsen13} as well as open clusters \citep[e.g.][]{Cargile14,Rebull16}. 
LS models the observed variability as a sine wave with a given period. As discussed in \citet{VanderPlas17}, LS is the optimal statistic for fitting a sinusoid to data, but this is not the same as being optimally suited to finding the period of a generic sinusoidal-like signal, such as photometric modulation arising from the rotation of spotted stars.
For a light curve displaying photometric modulation, LS determines the best-fitting sinusoid, which makes the implicit assumption that both the modulation period, and its phase shape (i.e. shape within a given rotation cycle), are constant in time. This is usually an acceptable assumption for low-mass stars with stable modulation patterns, but not for many solar-type stars, which display evolving signals (see \S \ref{sec:across_sample} and \ref{sec:morph} for further discussion).

\subsection{Generalised Autocorrelation (G-ACF)}
\label{sec:GACF}

Autocorrelation, i.e. correlating a dataset with itself, is a well-tested `model-free' approach to estimating rotation periods from stellar light curves \citep[see e.g.][for application to Kepler data]{McQuillan14}. Periodogram methods generally assume a sinusoidal basis function, which can lead to incorrect solutions when presented with non-sinusoidal signals. The model-free autocorrelation function (ACF), therefore, is applicable to all light curves, irrespective of the shape and evolution of the variability signal. 

Traditional autocorrelation is limited to regularly sampled time series (typically space-based monitoring data), which has been a limiting factor in its wider application to ground-based (and other non-regularly sampled) datasets. Here, we use a new generalised autocorrelation function (\gacf), which is a generalised version of the standard ACF that is applicable to both regularly and irregularly sampled time series. The algorithm is introduced in Kreutzer et al. (\emph{in prep.}) and described in detail in Briegal et al. (\emph{in prep.}), with specific application to extracting stellar rotation periods for field stars with \ngts. We therefore give only a brief description of the algorithm below and refer the interested reader to these publications.

Performing ACF on irregularly sampled data is possible if we: 1. generalise the ACF `lag' term from an integer multiple of the sampling constant to a real parameter, and 2. define selection and weight functions to decide how to identify and interpret correlations between data points whose timestamps do not perfectly align for a given lag (time shift). For each data point, and at each lag, we choose the closest un-shifted data point in time to correlate with, and weight the correlation between each pair of data points by a function that is inversely proportional to their time difference.

We estimate rotation periods from the ACF by calculating a two-stage Fast Fourier Transform (FFT). First, we calculate the FFT of the whole ACF to identify approximate periods, and then refine these by using only the section of the ACF within 5 times the identified period. Calculating FFT on the ACF is more appropriate than on the light curves themselves, as the ACF is more sinusoidal than the stellar modulation patterns in most cases. We refer the reader to Briegal et al. (\emph{in prep.}) for further details on this procedure. In the current work, the primary reason for this two-stage process was to account for evolution present in the rotation signals of solar-type stars (see \S \ref{sec:P_comp} for further discussion).

\subsection{Gaussian process regression}
\label{sec:GPs}

It has been demonstrated that a Gaussian Process (GP) can be used as a descriptive model to measure stellar rotation periods \citep{Angus18}. A GP is a model for the covariance between data points or, in other words, the autocorrelation of the time series. This means that we can use a GP with a quasiperiodic covariance and interpret the parameters of that model as physical properties of the time series. We extend the framework presented by \citet{Angus18} to include scalable computation of the GP model using the \textsl{celerite} algorithm \citep{Foreman-Mackey17} and we improve the runtime of the inference procedure by taking advantage of a scalable method for computing the gradient of \textsl{celerite} models with respect to their parameters \citep{Foreman-Mackey18}. This efficient calculation of gradients enables the use of the efficient No U-Turn Sampling \citep[NUTS;][]{Hoffman14} method for posterior inference. This fitting procedure is implemented as part of the \textsl{exoplanet} project \citep{exoplanet} and is built on top of \textsl{Theano} \citep{theano} for efficient model evaluation, \textsl{PyMC3} \citep{pymc3} for inference, and \textsl{AstroPy} \citep{astropy1,astropy2} for data manipulation.

The kernel function that we use to model the covariance caused by stellar variability is a mixture of three stochastically-driven damped simple harmonic oscillators (SHOs). This function is described in more detail by \citet{Foreman-Mackey17}, but each oscillator introduces one term into the description of the kernel and each term $k$ has the power spectrum 
\begin{equation}
S_k(\omega) = \sqrt{\frac{2}{\pi}} \frac{S_{0,k}\,{\omega_{0,k}}^4}
{{({\omega}^2 - {\omega_{0,k}}^2)}^2 + {\omega_{0,k}}^2\,{\omega}^2 / {Q_k}^2}
\end{equation}
where $\omega$ is the angular frequency, $S_{0,k}$ is the amplitude of the oscillation, $\omega_{0,k}$ is the undamped frequency, and $Q_k$ is the quality factor.

We fix $Q_1 = 1/\sqrt{2}$ so that the first term can capture any non-periodic covariance in the time series \citep{Foreman-Mackey17} and the other two terms are constrained to have frequencies that differ by a factor of two. Specifically, we define the parameters of the second and third terms as follows
\begin{eqnarray}
Q_2 &=& Q_3 + \Delta Q \\
\omega_{0,2} &=& \frac{4\,\pi\,Q_2}{P_\mathrm{rot}\,\sqrt{4\,{Q_2}^2-1}} \\
S_{0,2} &=& \frac{A}{\omega_{0,2}\,Q_2} \\
\omega_{0,3} &=& \frac{8\,\pi\,Q_3}{P_\mathrm{rot}\,\sqrt{4\,{Q_3}^2-1}} \\
S_{0,3} &=& \frac{f\,A}{\omega_{0,3}\,Q_3}
\end{eqnarray}
parameterised by a rotation period $P_\mathrm{rot}$, the quality factor of the third term $Q_3 > 1/2$, the difference between the quality factor of the second and third terms $\Delta Q > 0$, the amplitude of the base harmonic $A > 0$, and the fractional amplitude of the third term $0 < f < 1$.

We initialise the sampler using the rotation period estimate from either LS or \gacf\footnote{We note that the exact initial guess is not important as long as it is reasonably close to the actual rotation period.}, perform an initial maximum a posteriori (MAP) fit, identify 3$\sigma$ outliers from this initial solution, and then mask the outliers and refit. Masking these outliers was designed primarily to remove flares, but also other non-rotation phenomena such as eclipses. For most stars, around 10 data points were flagged as outliers and masked, although flaring stars typically had more. We then run four independent Markov chains for 2000 steps each to tune the mass matrix of the proposal, followed by four production chains of 2000 steps each. This procedure takes about ten minutes to run using two CPUs and generally results in several thousand effective samples of the rotation period that we use to approximate the posterior probability density.

\subsection{Comparison between LS, \gacf\ and GPs}
\label{sec:P_comp}

\subsubsection{Assumptions underlying the three methods}
\label{sec:assumptions}

The three methods differ in their assumptions and flexibility, which can be seen in their predictions for the rotation periods of stars displaying different light curve morphologies. Figure \ref{fig:P_agree} shows two examples of light curves that display regular modulation patterns, which do not significantly evolve during the 200 day light curves. LS, \gacf\ and GPs all typically predict consistent rotation periods for such stars. Figure \ref{fig:P_diff} shows two examples of light curves whose modulation patterns evolve significantly during the 200 day \ngts\ observations. For these two stars, GPs and \gacf\ agree to within 1$\sigma$ but the LS period is discrepant by $>$5$\sigma$ (compared to the GP posterior period distribution). The light curves and period results for all stars with detected rotation periods are given in the supplementary material in the online journal. The two sets of examples in Figures \ref{fig:P_agree} and \ref{fig:P_diff} highlight some of the assumptions underlying the three methods:
\begin{itemize}
\item LS is essentially a rigid sine-wave model and is therefore best-suited to light curves whose modulation patterns are purely sinusoidal and do not evolve throughout the observations. It is less well-suited to determining rotation periods from stellar light curves whose modulation signal (primarily the phase shape) evolves appreciably during the observations, for example from evolving active regions and/or differential rotation.

\item \gacf\ on the other hand, as the data itself is the model, is suitable for extracting periods from light curves irrespective of the modulation shape and evolutionary timescale. Complications from this method arise due to the diurnal nature of the observations, which give rise to a low-level 1-day signal imparted on the ACF that can subtly modify the exact shape of ACF peaks and hence the estimated rotation periods.

\item GPs lie somewhere between LS and \gacf\ in terms of their assumptions. They form a class of model, akin to LS, which depends on both the covariance kernel chosen and the covariance properties of the data being analysed. This drawing of information from the data itself shares some similarity with the principles underlying the \gacf\ method. What sets the GP method apart, in the current context, is that by drawing covariance information from the data and interpreting it through a kernel to generate a model, the GP has predictive power. This predictive power can be seen between individual nights and before and after the observations (see the second-from-top panels in Figures \ref{fig:P_agree} and \ref{fig:P_diff}). Furthermore, based on tests where subsections of light curves were masked, the GP models were able to adequately predict the stellar modulation patterns over a few-to-several rotation periods, depending on the level of evolution present in the light curve. 
\end{itemize}
Underlying all three methods is a general assumption that the light curve modulation patterns have a single underlying period. Multiple periods can be detected, however: in LS and \gacf, from two or more unrelated peaks, and for GPs by double-peaked period distributions (if sufficiently close that the MCMC explores that part of parameter space). Of course, to correctly estimate individual periods from a single light curve when multiple periodic signals are present, either due to differential rotation or the system being a multiple star system, a composite model needs to be applied, which is not the case here. Therefore, we make no attempt to try and characterise differential rotation in our light curves or characterise the rotation periods of individual stars in identified multiple star systems. We refer the reader to \citet{Gillen17} as an example of such an effort, but leave this to future work here.

\subsubsection{Comparison across the \blanco\ sample}
\label{sec:across_sample}

Of the \nstars\ Blanco 1 stars observed by NGTS, we detected rotation periods in \nProt\ stars. For \nProtsame\ of these, all three methods detected the same modulation signal, but for the remaining \nProtdiff\ stars either aliases or different signals were preferred by one or more methods.

\begin{figure}
	\includegraphics[width=\columnwidth]{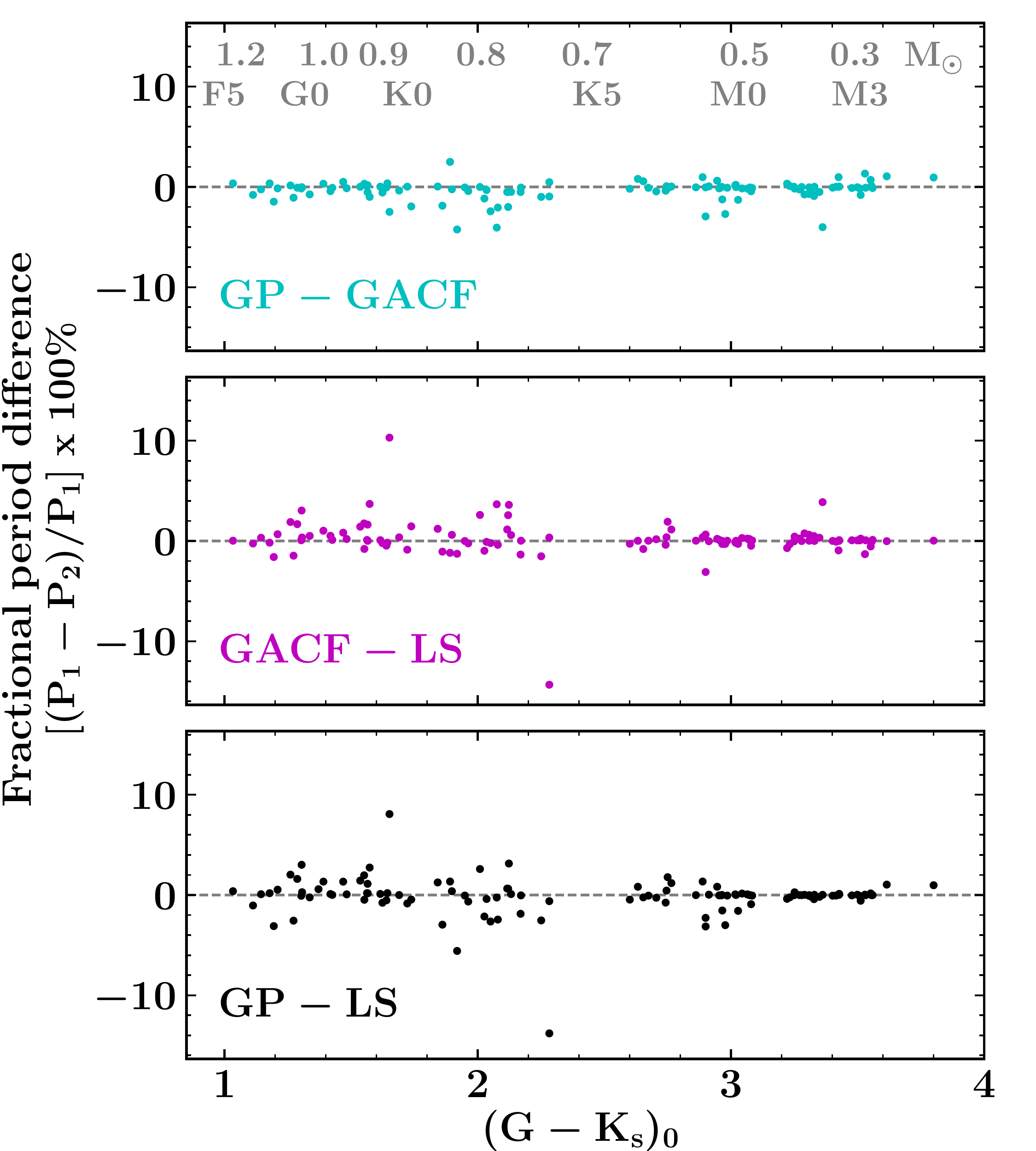}
    \caption{Comparison of the periods extracted for \blanco\ stars using GP, \gacf\ and LS methods. We compare the period differences between each method (defined as $[(P_{1}-P_{2})/P_{1}] \times 100 \%$) as a function of dereddened \gmk\ colour. For this comparison, we use \nProtsame\ stars where each method detected the same rotation signal. While the agreement is good for most stars, there are exceptions where different methods disagree by up to $\sim$15\%. The agreement is better for the lower mass stars (\igmk\,$\gtrsim$\,2.5), which typically show stable modulation signals. Solar-type stars (\igmk\,$\lesssim$\,2.5) display greater evolution in their light curves, which causes a larger scatter in the period estimates between the three methods. Overall, the GP and \gacf\ methods (cyan) agree best, most notably for the evolving solar-type members, followed by \gacf\ and LS (magenta) and then GP and LS (black).}
    \label{fig:P_comp}
\end{figure}

In Figure \ref{fig:P_comp} we compare the agreement between the periods extracted by our three methods using their fractional period differences, i.e. $[(P_{1}-P_{2})/P_{1}] \times 100 \%$ as a function of \igmk\ colour. We opted to compare against \igmk\ colour rather than rotation period because it is a more fundamental parameter (being a proxy for mass) and better separates different light curve morphologies (see \S \ref{sec:morph}). Essentially, mid-F to mid-K stars (\igmk\,$\lesssim$\,2.5) typically show evolution in their modulation patterns (both in amplitude and phase shape) whereas late-K and especially M stars (\igmk\,$\gtrsim$\,2.5) generally display more stable sinusoidal modulation. The evolutionary nature of the light curves is the main cause of disagreement between methods. We note that the faintest stars in our sample sometimes contained residual moon variations arising from incomplete background correction, which we fit and removed as described in Appendix \ref{sec:appendix}. For these stars, the three methods typically agree well.

The GP and \gacf\ periods (cyan) agree best, as they are the two most flexible methods, whose agreement is consistently within $\sim$2--3\% across the whole \blanco\ sample. Both \gacf\ vs. LS (magenta) and GP vs. LS (black) show the same general trend: the agreement is good for late-K and M stars (\igmk\,$\gtrsim$\,2.5), whose modulation patterns are stable and sinusoidal, but are noticeably worse for the mid-F to mid-K stars (\igmk\,$\lesssim$\,2.5), which display evolving modulation patterns. The worse agreement for the mid-F to mid-K stars is primarily due to the rigid nature of the LS algorithm, which does not allow for any evolution, and essentially finds the period that folds the data closest to a sine wave. GPs and \gacf, on the other hand, are flexible enough to account for the evolution seen in the \blanco\ light curves. 

We find that GPs and \gacf\ perform best across the \blanco\ sample (F5 to M3 stars). It is worth noting that GPs require an initial guess for the rotation period whereas \gacf\ does not. We therefore see \gacf\ as an efficient method for detecting rotation periods from large samples of stars that display a range of modulation morphologies, and GPs as a powerful tool to refine period estimates and to better understand modulation signals.

We visually inspected the GP, LS and \gacf\ results for all light curves. We found that the GP method gave the most reliable rotation periods across the full sample and hence selected the GP periods for all but four stars (where we favoured either the \gacf\ or LS period). These four stars possess short ($\lesssim 1$ day) periods and stable modulation. We note that the difference between the \gacf\ or LS period and the GP period for these stars was $\lesssim$\,15 mins (0.01 days), so our choosing the \gacf\ or LS periods for these stars was to present the best periods we could, rather than the GP period being incorrect per se. As we did not compute errors for the \gacf\ and LS periods, we note that these four stars do not have errors associated with their favoured periods but, based on the range of periods resulting in well-defined phase-folded modulation signals, they should all have errors on the order of $\sim$0.02 days or less. Table \ref{tab:periodic} reports period information for all stars detected to be periodic, and includes the GP, \gacf\ and LS periods for all stars, along with the method adopted. Additional information, including positions, magnitudes, colours, estimated spectral types and detected multiplicity (see \S \ref{sec:identify_multiples}), is also given. Table \ref{tab:non_periodic} reports the same set of additional information for stars not detected to be periodic.

\begin{landscape}

\begin{table}
  \caption{Identification, photometric, multiplicity and period information for periodic \blanco\ stars. The full table is available in machine-readable format from the online journal.} 
  \label{tab:periodic}
  \resizebox{\linewidth}{!}{%
  \begin{tabular}{rccccccccclcccccccccc}
    \hline
    \hline
    \noalign{\smallskip}

    NGTS ID  &      Gaia ID  &      RA  &       Dec  &       NGTS   &      G  &      G$_{\rm BP}$ &      G$_{\rm RP}$ &      K$_{\rm s}$  &     G$-$K$_{\rm s}$  &      SpT   &      multiple\,*  &     Amp  &      Amp  &      P$_{\rm GP}$   &      P$_{\rm GP}^{\rm \, uerr}$   &      P$_{\rm GP}^{\rm \, lerr}$   &      P$_{\rm LS}$   &      P$_{\rm GACF}$   &      Method  &      P$_{\rm adopted}$   \\
    &  & (J2000) & (J2000) & mag &  &  &  &  &  &  &  & (data) & (GP) &  &  &  &  &  &  &  \\

    \noalign{\smallskip}
    \hline
    \noalign{\smallskip}
    
    626  &  2321422436144254464  &  00:06:53.70  &  -28:31:08.26  &  11.87  &  12.25  &  12.64  &  11.71  &  10.66   &  1.59  &  G8    &  --   &  17.78   &  15.33   &  5.37994   &  0.02861   &  0.02901   &  5.36998   &  5.37530  &  GP     &  5.37994 \\
    
    1221  &  2333096436428744192  &  00:01:08.42  &  -28:36:56.81  &  11.15  &  11.49  &  11.82  &  11.03  &  10.20   &  1.29  &  G0    &  --   &  15.95   &  20.85   &  3.03852   &  0.01165   &  0.01176   &  2.97678   &  3.03410  &  GP     &  3.03852  \\
    
   1283  &  2333061733093062784  &  00:03:38.49  &  -28:37:25.34  &  11.71  &  12.07  &  12.44  &  11.54  &  10.57   &  1.49  &  G5    &  --   &  34.00   &  34.08   &  5.02679   &  0.02690   &  0.02528   &  4.95986   &  5.00120  &  GP     &  5.02679  \\
   
   2442  &  2321365197116041984  &  00:09:14.70  &  -28:47:13.64  &  15.44  &  16.01  &  17.22  &  14.93  &  12.57   &  3.44  &  M2.5  &  --   &  36.83   &  21.43   &  2.11056   &  0.00080   &  0.00080   &  2.11170   &  2.11025  &  GP     &  2.11056  \\
   
   2895  &  2333042143747087616  &  00:04:02.37  &  -28:51:22.38  &  15.87  &  16.39  &  17.70  &  15.25  &  12.83   &  3.56  &  M3    &  --   &  79.78   &  65.67   &  1.64759   &  0.00035   &  0.00033   &  1.64787   &  1.64900  &  GP     &  1.64759  \\
   
   3201  &  2321349700873987712  &  00:07:29.91  &  -28:53:41.83  &  14.84  &  15.45  &  16.61  &  14.37  &  12.09  &   3.36  &  M2  &    c  &    42.15  &   38.92  &   0.32369  &   0.00005  &   0.00004  &   0.32363  &   0.32364  &  GP  &     0.32369 \\

   ...  &  ...  &  ...  &  ...  &  ...  &  ...  &  ...  &  ...  &  ...  &  ...  &  ...  &  ...  &  ...  &  ...  &  ...  &  ...  &  ...  &  ...  &  ...  &  ...  &  ... \\
       
    \noalign{\smallskip}
    \hline
  \end{tabular} }%
  \begin{list}{}{}  
  \item{$^{*}$ c = CMD and r = RV. For example, ``cr'' would indicate a system that was highlighted as a likely multiple system by both methods.}
  \end{list} 
\end{table}

~\\

\begin{table}
  \centering
  \caption{Identification, photometric and multiplicity information for \blanco\ stars without a detected period. The full table is available in machine-readable format from the online journal.} 
  \label{tab:non_periodic}
  \begin{tabular}{cccccccccclc}
    \hline
    \hline
    \noalign{\smallskip}
    NGTS ID  &  Gaia ID  &  RA  &   Dec  &   NGTS  &  G  &  \bp\ &  \rp\ & K$_{\rm s}$  & \gmk\  &  SpT  &  multiple$^{*}$  \\
      &    &  (J2000)  &  (J2000)  &  mag  &    &    &    &  &  &  &  \\
    \noalign{\smallskip}
    \hline
    \noalign{\smallskip}
    
  3284  &  2321395090088371456  &  00:07:13.73  &  -28:54:38.94  &  15.22  &  15.82  &  17.21  &  14.67  &  12.22  &  3.60  &  M3    &  c  \\
  3591  &  2333019328880804096  &  00:03:27.55  &  -28:57:38.55  &  14.64  &  15.23  &  16.33  &  14.16  &  11.92  &  3.31  &  M1.5  &  c  \\
  4258  &  2320971606313124352  &  00:06:51.14  &  -29:02:58.37  &  16.00  &  16.57  &  17.92  &  15.43  &  13.02  &  3.55  &  M3    &  -- \\
  5249  &  2321010329738234112  &  00:06:09.05  &  -29:09:10.48  &  8.41  &  7.89  &  7.89  &  7.93  &  8.00  &  -0.10  &  B9  &  --  \\
  6610  &  2320945699070402816  &  00:06:53.50  &  -29:21:13.25  &  14.26  &  14.86  &  15.75  &  13.95  &  11.97  &  2.89  &  K8  &  -- \\
  6761  &  2320950333339180544  &  00:07:25.36  &  -29:22:25.12  &  15.38  &  16.05  &  17.33  &  14.82  &  12.43  &   3.62  &  M3  &    c  \\
  ...  &  ...  &  ...  &  ...  &  ...  &  ...  &  ...  &  ...  &  ...  &  ...  &  ...  &  ...  \\
      
    \noalign{\smallskip}
    \hline
  \end{tabular}
  \begin{list}{}{}  
  \item{$^{*}$ c = CMD and r = RV. For example, ``cr'' would indicate a system that was highlighted as a likely multiple system by both methods.}
  \end{list}
\end{table}

\end{landscape}

\subsection{Comparison with literature rotation periods for \blanco}
\label{sec:P_comp_lit}

\begin{figure}
	\includegraphics[width=\columnwidth]{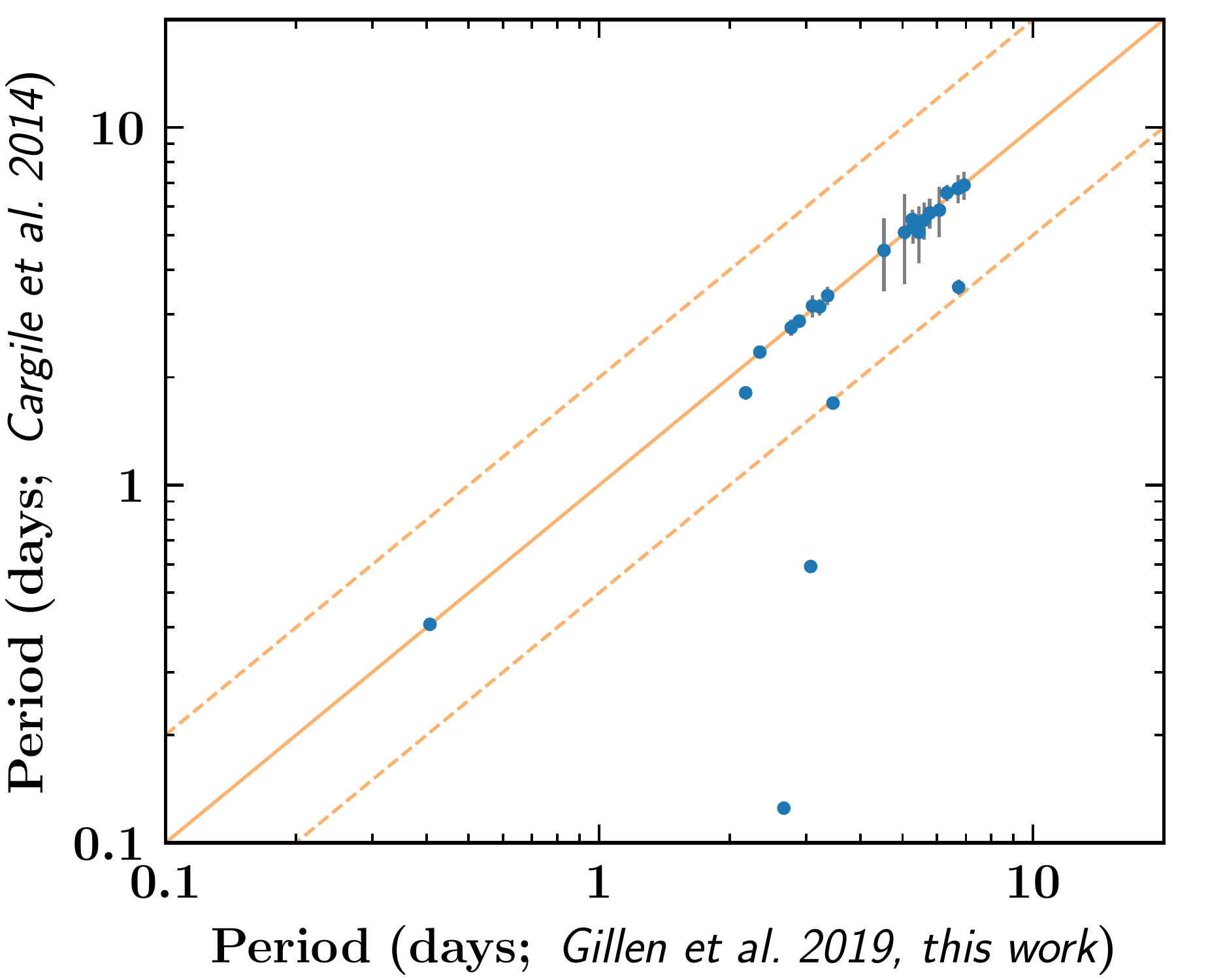}
    \caption{Comparison between our rotation periods and those of \citet{Cargile14} for 23 stars with detected periods in both surveys. Of these, 18 have periods that agree to within 10\% or better. The solid and two dashed orange lines show the 1:1 period match, and the 2:1 and 1:2 harmonics. The period uncertainties reported in this work are typically the size of the points or smaller.}
    \label{fig:P_comp_C14}
\end{figure}

We compare our rotation periods to literature values from \citet[][hereafter C14]{Cargile14} as a further sanity check for the novel GP and \gacf\ techniques presented here. C14 estimated rotation periods for bright \blanco\ members using KELT-South light curves, which comprised 43 nights of data spread over 90 nights. Figure \ref{fig:P_comp_C14} shows the match between our rotation periods and theirs for 23 stars that have detected periods in both surveys and are DR2-confirmed members of the cluster. 

18 of the 23 stars (78\%) have rotation periods that agree to 10\% or better, with 17 of these agreeing to within their 1$\sigma$ uncertainties. Of the other five stars: one (NGTS ID 12805\,\footnote{See online supplementary figures.}) has a period that agrees to within 20\%, which suggests that the same rotation signal is being probed, but the detected period is probably affected by correlated noise in one or both datasets; two stars (NGTS IDs 14186 and 10246\,\colblue{$^{5}$}) lie on the 1:2 harmonic, with C14 reporting periods that are half of ours; one star (NGTS ID 8749\,\colblue{$^{5}$}) has a period in C14 that lines up with an alias of our period in the LS periodogram; and the last outlier (NGTS ID 12641\,\colblue{$^{5}$}) has a very short period in C14, which is indicative of a binary system, but we do not identify it as such here and find its rotation period places it on the well-defined single star sequence for its colour. From comparison of our light curves with those in C14, and the fact that all three of our methods agree for the stars not lying on the 1:1 relation, we favour our rotation periods for all stars.

\section{Identifying multiple stars}
\label{sec:identify_multiples}

We identify binary and higher order multiple stars using two complementary approaches: 1. fitting the single star cluster sequence in colour-magnitude space and identifying stars lying above this trend; and 2. cross-matching with literature radial velocity (RV) surveys. We describe each method in \S \ref{sec:CMD_fitting} and \ref{sec:RV_surveys} below.

\subsection{CMD fitting}
\label{sec:CMD_fitting}

\begin{figure*}
	\includegraphics[width=\columnwidth]{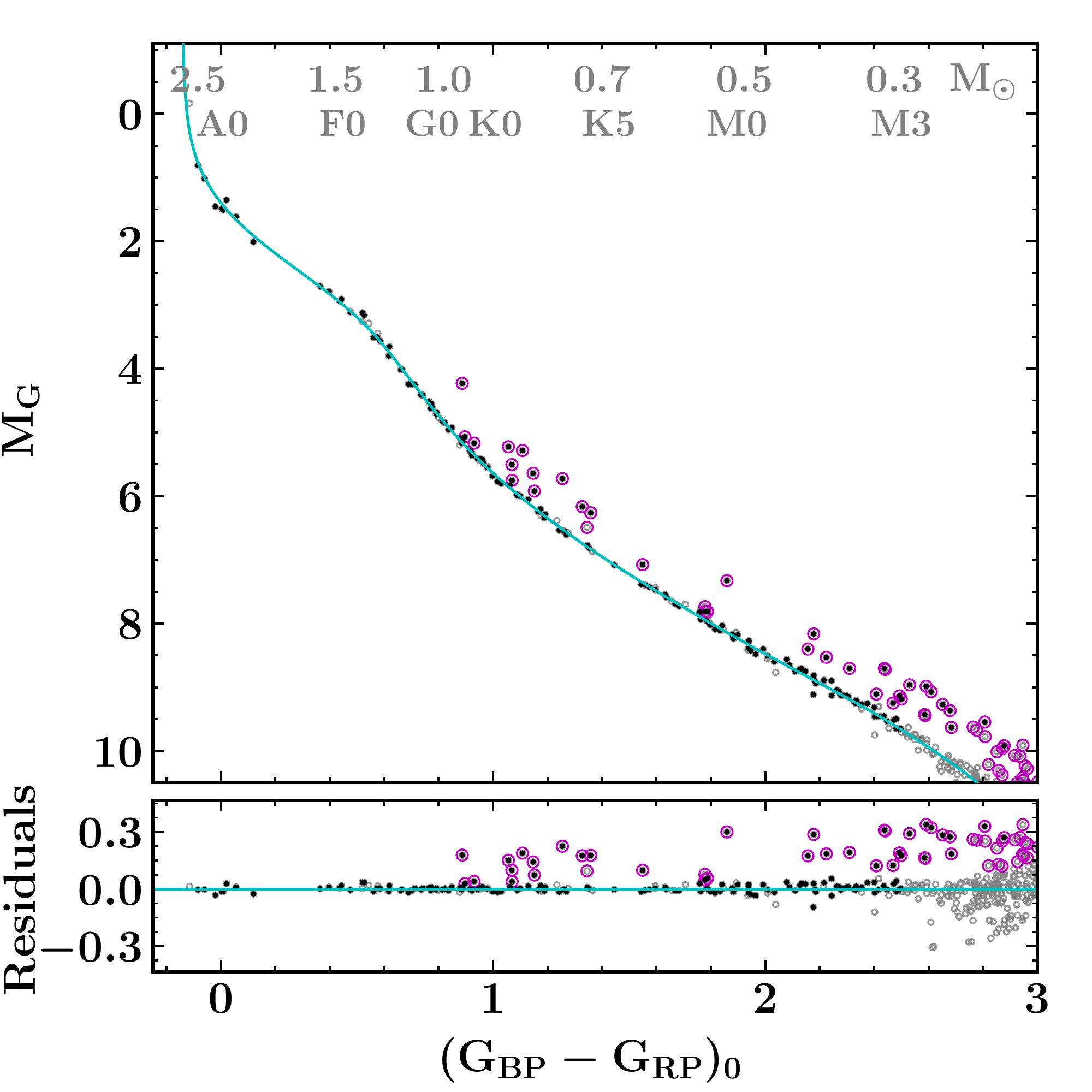}
	\includegraphics[width=\columnwidth]{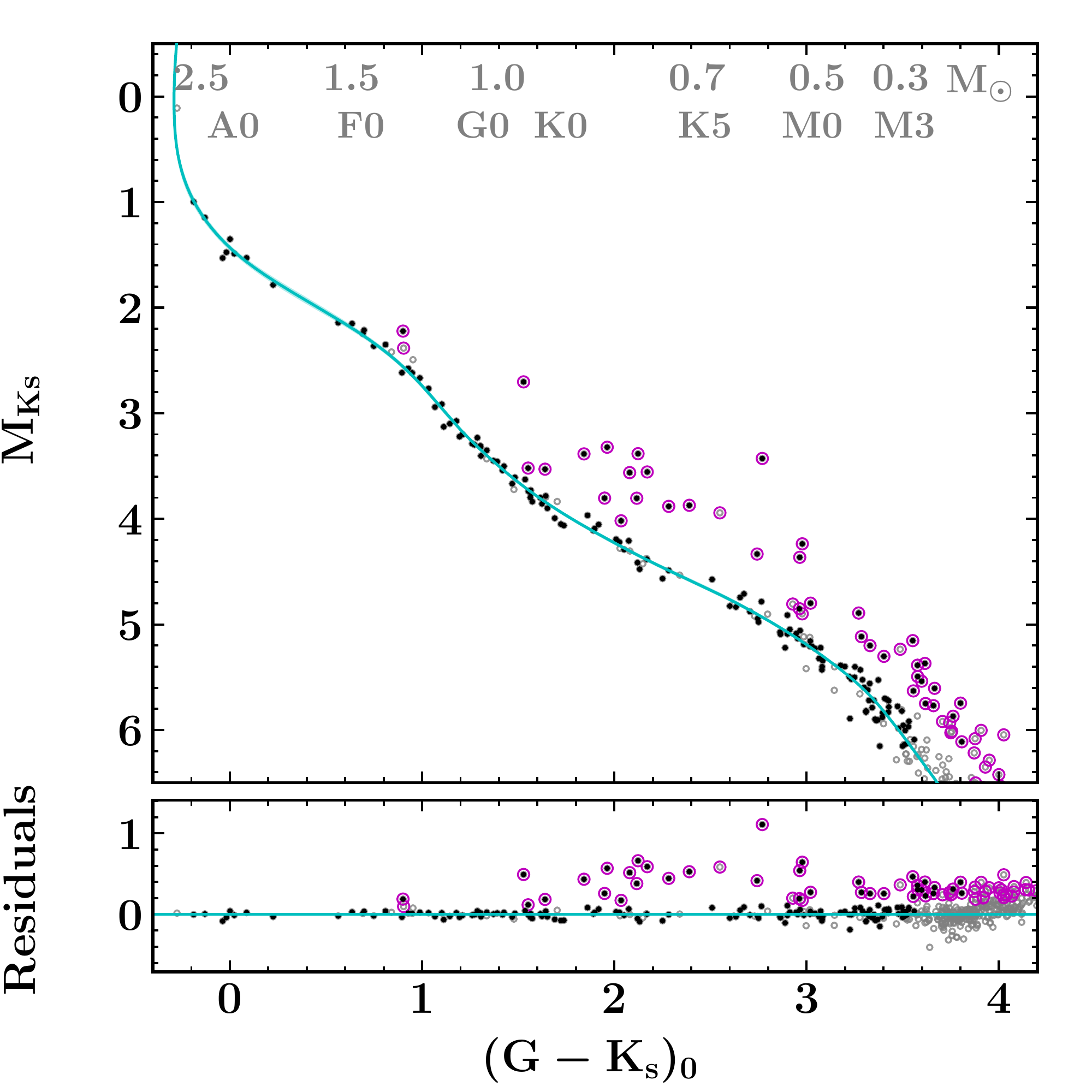}
    \caption{Colour-magnitude diagrams (CMDs) of \blanco. \emph{Left:} Absolute G vs. dereddened \bprp\ CMD of all \blanco\ stars (open grey circles) and those with \ngts\ light curves highlighted (filled black circles). The single star locus has been estimated by iteratively fitting the cluster sequence using a Gaussian process (cyan). Stars that are 3$\sigma$ outliers are circled in magenta and are likely multiple star systems. Below are the residuals of the fit as a function of colour, where `residual' here means the smallest linear distance from the GP model rather than vertical magnitude displacement above the single star sequence. \emph{Right}: same for absolute \ks\ vs. dereddened \gmk\ CMD.}
    \label{fig:CMD_fit}
\end{figure*}

We created four colour-magnitude diagrams to help us identify binaries with different mass ratios. These were \gaia\ \mg\ vs. \ibprp, \mg\ vs. \igmk, 2MASS \mk\ vs. \ibprp\ and \mk\ vs. \igmk. From the optical \mg\ vs. \ibprp\ CMD, relatively equal mass binaries are typically easy to identify as their colour remains roughly constant but their magnitude increases, with an equal mass binary lying 0.75 mag above the single star sequence. Low-mass ratio binaries are harder to detect in such an optical CMD, however. Taking the example of a G star with an M-dwarf companion, the M-dwarf will not affect the optical G magnitude much, and will only shift the BP-RP colour slightly redder, as the M-dwarf contributes more to the red-optical flux than the blue. Such a small colour shift is difficult to detect even with the precision of the Gaia data. An \mk\ vs. \igmk\ CMD is better suited for identifying low mass ratio binaries, as the presence of a low mass companion will more strongly contribute to the infrared \ks\ band and therefore shift the binary in both \ks\ magnitude and \gmk\ colour. By using four CMDs, we can track the relative positions of each star in the four planes, and gain a better handle on their likelihood of being a multiple star system.

With the precision of \gaia\ data, stellar evolution models struggle to fit the exact shape of cluster CMDs at a given age and metallicity (see e.g. \colblue{B18}); this motivated us to use a flexible non-parametric model. For each CMD, we iteratively fit the whole cluster sequence using a GP with a stochastically-driven damped simple harmonic oscillator (SHO) kernel, as implemented in {\tt Celerite} and {\tt exoplanet}, and fixed the quality factor at $Q=1/2$, which approximates the well-known Matern-3/2 kernel \citep{Rasmussen06}. At each step, we perform a running median filter on the residuals of the fit and reject 3$\sigma$ outliers above the GP model in the next iteration, with this process typically converging towards a maximum a posteriori (MAP) fit to the single star cluster sequence within $\sim$5 iterations.

Figure \ref{fig:CMD_fit} shows the \mg\ vs. \ibprp\ and \mk\ vs. \igmk\ CMDs (left and right, respectively).While all four CMDs display a clear single star cluster sequence with multiple star outliers above the trend, the exact positions of individual multiple systems in each CMD can differ quite significantly due to their component mass-ratios.
We identify multiple star systems as those which lie at least 3$\sigma$ above the single star GP sequence in any one plane. In practice, if a system was an outlier in one plane it was typically an outlier in two or more planes. From the four CMDs, we identify \nphotbin\ multiple star systems from the \nstars\ stars with \ngts\ light curves.

\subsection{Cross-matching with literature radial velocity surveys}
\label{sec:RV_surveys}

We cross-matched our cluster sample with literature RV surveys, namely those of \citet{Mermilliod08,Mermilliod09} and \citet{Gonzalez09}, and identify seven SB1 and two SB2 binaries within our members that have \ngts\ light curves. These systems are identified in Tables \ref{tab:periodic} \& \ref{tab:non_periodic}. 
Three of the five periodic systems are also identified as binaries from our CMD analysis. 
In the following sections we use only the photometric multiples identified using our CMD analysis, and therefore do not consider the other two periodic systems (IDs 9992 and 11872) as multiples, although we do list them as such in Table \ref{tab:periodic}. This is for two reasons: 1. the literature RV surveys of \blanco\ are not complete in terms of either membership or spectral type, which makes the identified multiples hard to interpret statistically; and 2. in \S \ref{sec:B1_plei_comp} we compare to the \pleiades, for which we identify multiples using our CMD analysis only, and therefore use the same multiple star criteria for \blanco\ for fair comparison.


\section{Rotation in Blanco 1}
\label{sec:B1_rot}

\begin{figure*}
	\includegraphics[width=0.48\linewidth]{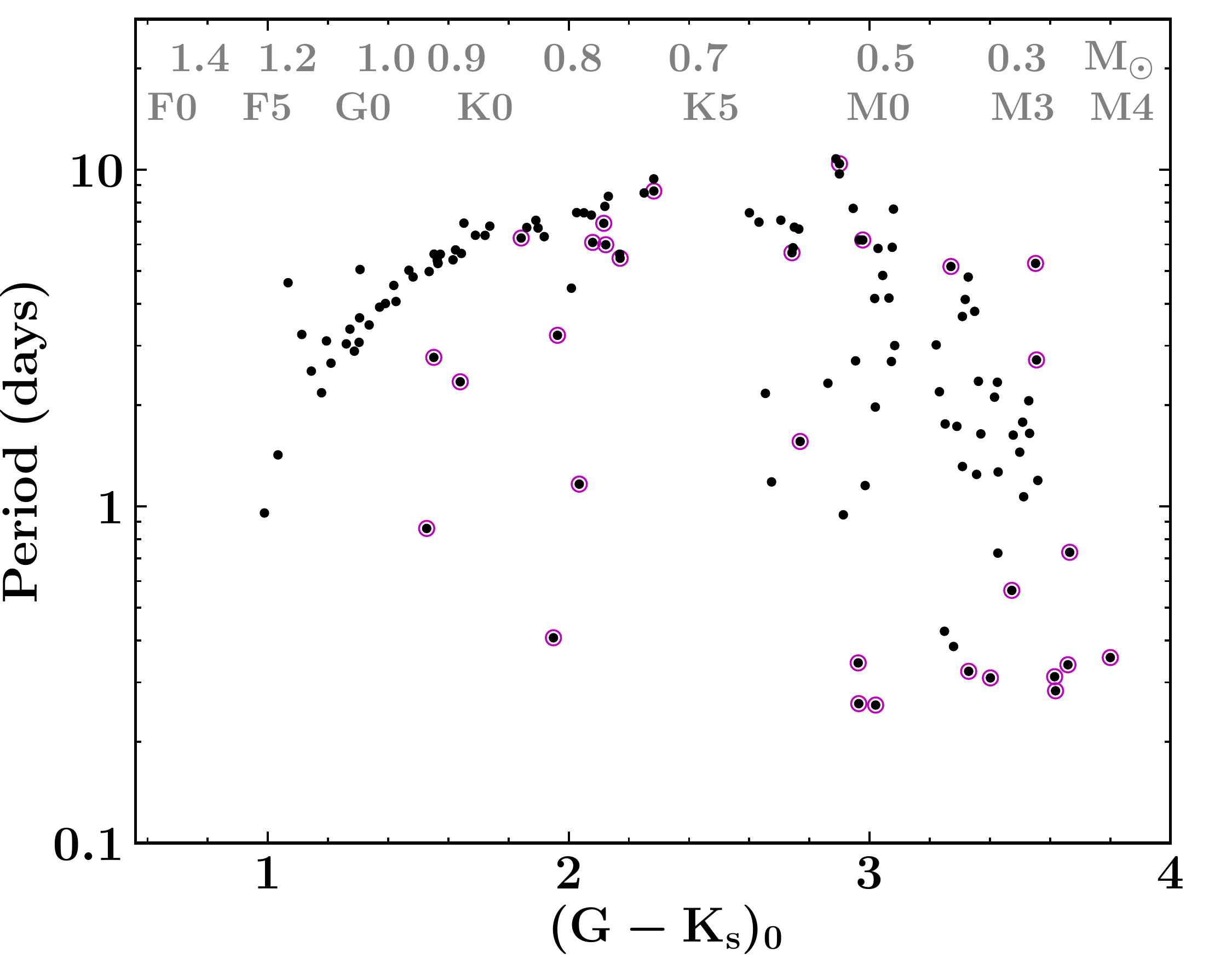}
	\includegraphics[width=0.48\linewidth]{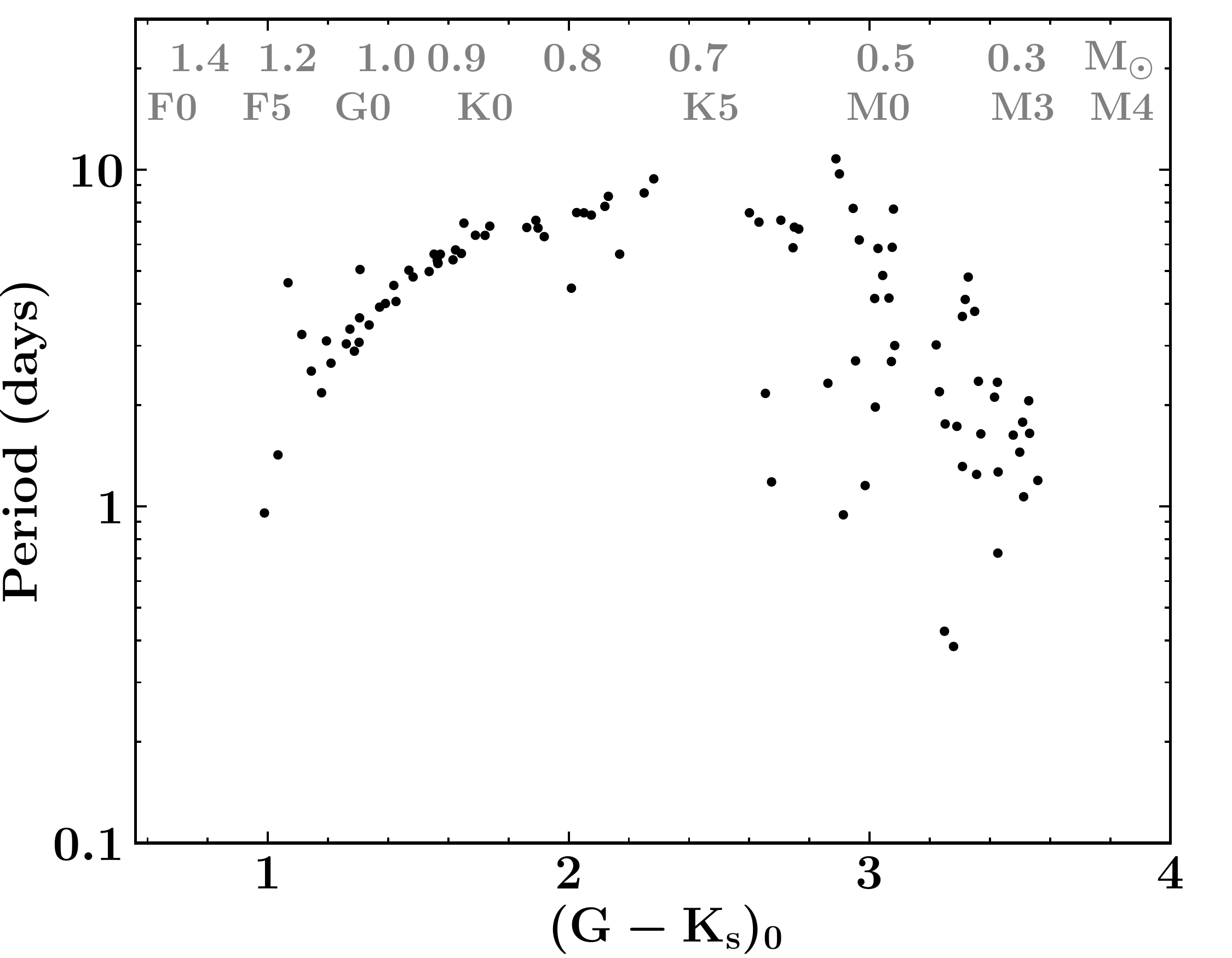}
    \caption{Rotation period vs. dereddened \gmk\ colour for stars in \blanco. \emph{Left}: all stars with detected rotation periods (black points), with our identified multiple stars circled in magenta. Stellar mass (\msun) and spectral type are indicated at the top. \emph{Right}: showing only the apparently single stars to highlight the clear rotation sequence between 1.1$<$\igmk$<$2.3 (1.2$\,\gtrsim M \gtrsim$\,0.75\,\msun). Mass-dependent angular momentum evolution is strongly imprinted in the \blanco\ sample.}
    \label{fig:P_col}
\end{figure*}

\subsection{Colour--period distribution}
\label{sec:col_P_dist}
\subsubsection{FGK stars with masses $0.7\lesssim M \lesssim 1.2$ \msun}
\label{sec:col_P_dist_FGK}

The rotation period distribution of stars in \blanco\ as a function of \igmk\ colour is shown in Figure \ref{fig:P_col}. In the left hand plot we show all stars for which we determined rotation periods and on the right we remove stars identified as likely multiple systems (see \S \ref{sec:identify_multiples}) to highlight the trend in the (apparently) single star population. For these single stars, there is a clear mass-dependence in the rotation distribution, which is especially evident for the mid-F to mid-K stars (1.0\,$<$\,\igmk\,$<$\,2.5, 1.2\,$\gtrsim M \gtrsim$\,0.7\,\msun), as they follow a tight rotation sequence between 2--10 days. Stars lying on this tight sequence are almost exclusively likely single stars, whereas those lying under the sequence at shorter rotation periods are almost exclusively likely multiple star systems.

The origin of this dichotomy is not well understood. All of the multiples identified here are photometric multiples, and hence are moderate-to-high mass ratio systems\footnote{We note that one of the two RV-detected binary systems, which was not also a photometric multiple, sits on the well-defined sequence of apparently single stars, while the other sits above (possibly because the secondary component is responsible for the modulation signal).}. In the field, the mass ratio distribution for multiples shows a preference towards high mass ratio pairs and that these occur more frequently in relatively close configurations \citep{Raghavan10}. One might expect, therefore, that many of the photometric multiples identified here also have relatively close separations. This may be important because stars that form in close-separation ($\lesssim$100 AU) multiples are thought to have their circumstellar disk lifetimes truncated due to the presence of their close companions \citep[e.g.][]{Patience02,Meibom07,Daemgen12,Daemgen13}. Reduced disk lifetimes imply shorter phases of magnetic disk-braking, and hence an earlier spin up towards the ZAMS that presumably results in faster (and perhaps more widespread) rotational velocities at the start of MS evolution compared to single stars of the same mass. Alternatively, the star formation process may deposit angular momentum differently in close multiple systems compared to single or wide multiples \citep{Larson03}, which may also result in different rotation period distributions on the ZAMS. In any case, surface rotational velocities decrease as stars evolve off the ZAMS due to angular momentum loss through magnetised stellar winds and redistribution within the stellar interior. If stars in close multiples do possess a different rotation period distribution compared to single stars on the ZAMS, it follows that this will persist during the early MS before all non-tidally locked stars eventually converge towards the same rotation period distribution. 
It remains unclear, on both observational and theoretical grounds, how long such convergence might take, but it does not appear to have occurred by the age of \blanco\ ($\sim$115 Myr), at least for G and K stars.
By the age of Praesepe (700--800 Myr), however, it seems as though essentially all non-tidally locked FGK stars, irrespective of their hierarchy, have converged to a well-defined rotation sequence \citep{Rebull17}. Further observations of clusters, specifically those with ages between $\sim$100--800 Myr, are needed to constrain this convergence timescale. It is worth noting that some of the \blanco\ multiples identified here may be very close-separation binaries (orbital periods less than $\sim$10--15 days), whose rotational evolution will be driven by tidal forces that act to synchronise the stellar rotation periods to that of the binary orbital period, and hence will never converge onto the single star rotation sequence. 
Based on very limited RV monitoring by \citet{Mermilliod09} of five of these multiples, however, this does not appear to be the case for all systems: while two are spectroscopic binaries, three appear to have essentially flat RVs to within their uncertainties. Further RVs are needed to confirm this tentative statement.

We highlight two stars (with \igmk\ colours of 2.0 and 2.3) that lie below the single star trend, but which do not appear to be photometric or spectroscopic multiples. We suggest that both of these systems are either: 1. multiples containing very low-mass companions, such that they are not identified by us as multiples; or 2. single stars whose angular momentum loss has been inhibited at some point during their evolution. We consider the first option as the more likely given the correlation between multiplicity and faster rotation in this mass range. 

Finally, we note that roughly half of $\sim$F6--K3 stars are expected to reside in binary or higher order systems \citep{Raghavan10}. We identify $\sim$20\,\% of the FGK stars as moderate-to-high mass ratio multiples, which suggests that some of the apparently single stars in our sample that lie on the well-defined rotation sequence, are likely low mass ratio multiples. Given this, we surmise that the presence of a low-mass companion alone is not sufficient to significantly affect the angular momentum evolution of the primary star. This is evidenced by one of the two RV-detected multiples that was not also a photometric multiple, which lies on the well-defined rotation sequence for FGK stars (although we note that the other such system sits above this sequence, possibly because the secondary component is responsible for the modulation signal).

\subsubsection{Late-K and M stars with masses $0.3\lesssim M \lesssim 0.6$ \msun}

For late-K and M stars (2.5\,$\lesssim$\,\igmk\,$\lesssim$\,3.6, 0.7\,$\gtrsim M \gtrsim$\,0.3\,\msun) there is no well-defined rotation sequence, with the apparently single stars possessing rotation periods ranging from P$<$1 day up to P$\sim$11 days. Within this colour range multiple star systems are spread throughout the single star population, although there is an accumulation of multiples at short (P$<$0.5 day) periods. It is likely that this accumulation is, at least partially, a selection effect, as the faintest stars in our sample are mostly multiple systems that are overly bright compared to single stars of the same mass.

\subsection{Light curve morphology}
\label{sec:morph}

\begin{figure}
    \includegraphics[width=\columnwidth]{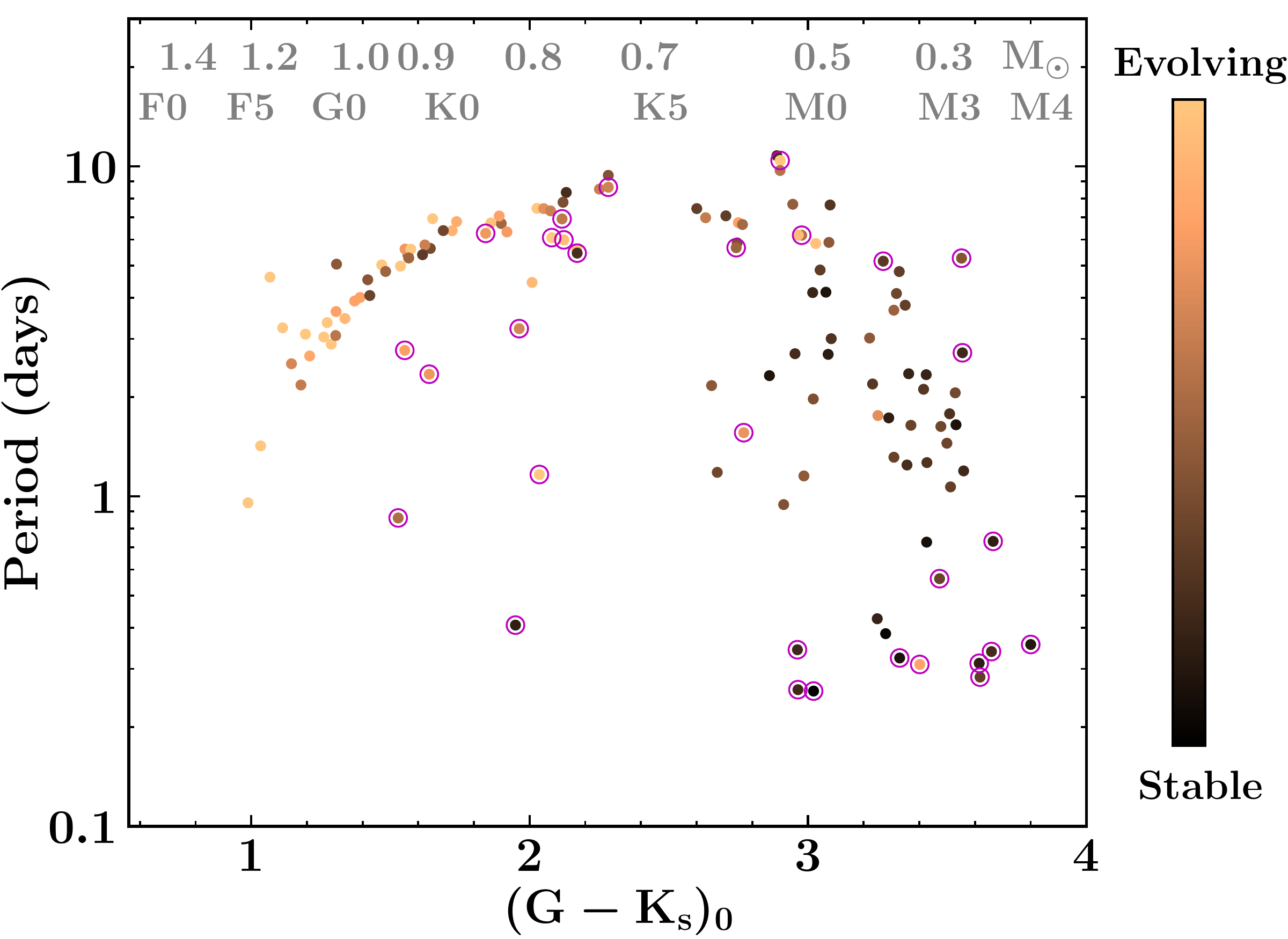}
    \caption{Rotation period vs. dereddened \gmk\ colour for stars in \blanco\ coloured by the level of evolution in their light curve modulation patterns. There is a clear mass dependence to the light curve morphology evolution, with mid-F to mid-K stars displaying predominantly evolving modulation patterns and M stars showing typically stable modulation over the 200 day \ngts\ light curves.}
    \label{fig:P_col_lcm}
\end{figure}

\begin{figure*}
\centering
  \includegraphics[width=\linewidth]{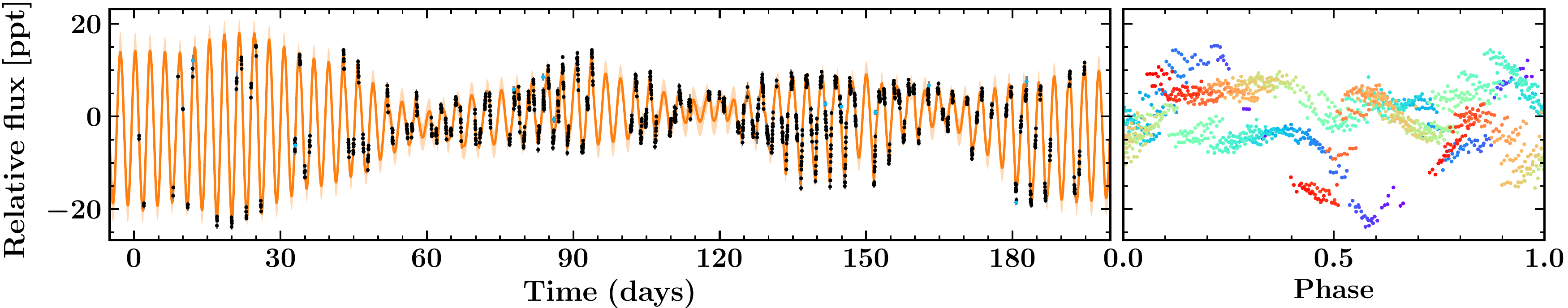}
  \includegraphics[width=\linewidth]{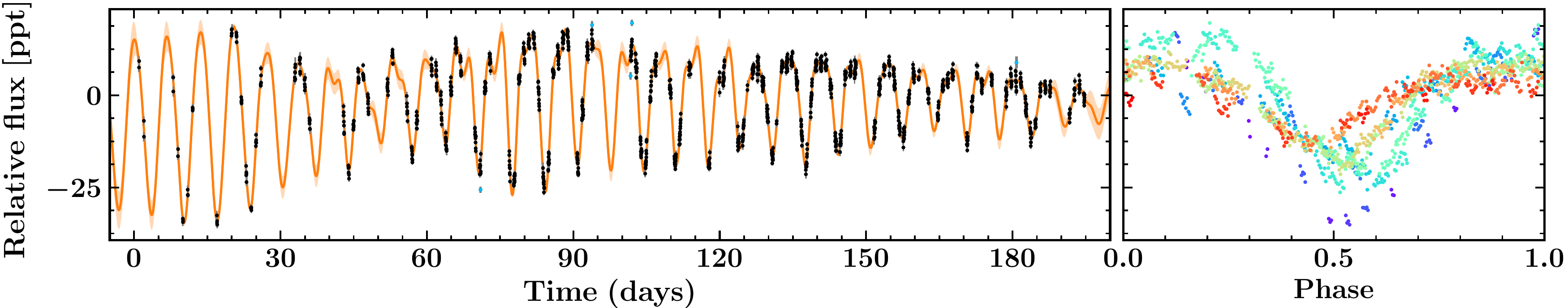}
  \includegraphics[width=\linewidth]{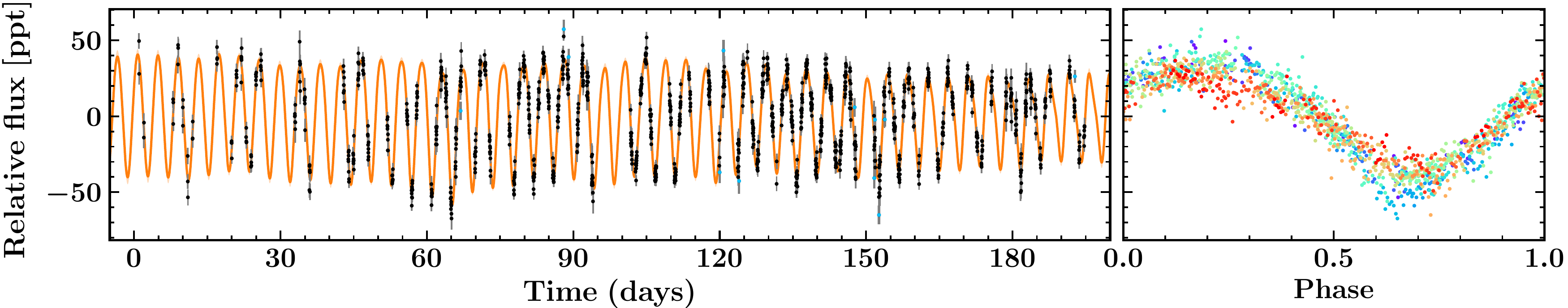}
\caption{Three example light curves displaying different levels of evolution. \emph{Top-to-bottom}: light curves showing strong evolution, moderate evolution, and stable modulation. In each case, the left panel shows the full light curve with the GP model fit, and the right panel the phase-folded light curve. The rainbow colour scheme represents individual modulation periods from the beginning (indigo) to the end (red) of the light curve.}
    \label{fig:lcm_examples}
\end{figure*}

\subsubsection{Evolving vs. stable modulation patterns}

The \ngts\ light curves offer an ideal window onto the evolution of young star ($\sim$115 Myr) modulation patterns over a well-sampled 200 day baseline. We classified the \ngts\ light curves based on two metrics: 1. the spread in the dispersion in the phase-folded light curve and 2. the self-similarity of the modulation pattern throughout the light curve. For the latter metric, we used running windows of 3--5 rotation periods, folded the data within each window on the global rotation period, and compared the difference in the flux measurements at similar phases throughout the light curve. Both metrics are probing the self-similarity of the data throughout the \ngts\ light curves, but we found that they were more sensitive to different aspects of the evolution. Combining these two metrics, by adding them in quadrature, provided a reasonably good indication of the level of evolution within a given light curve. We show the results in Figure \ref{fig:P_col_lcm}, where we see a strong mass dependence in the light curve morphologies: 
the majority of mid-F to mid-K stars show moderate-to-significant evolution (i.e. varying amplitudes and phase shapes) whereas M star modulation patterns appear generally stable over 200 days, with a transition between these occurring at late-K spectral types. 
In Figure \ref{fig:lcm_examples} we show three example light curves that highlight the range in evolution present within the \blanco\ stars, from more-to-less evolution (top-to-bottom).

Physically, this difference may be due to a change in the dominant magnetic field morphology between mid-F to mid-K stars and M dwarfs at this age. 
It is interesting to note that the change from predominantly evolving modulation patterns to predominantly stable patterns occurs around the same mass as the change from a well-defined rotation sequence to a broad rotation period distribution (i.e. M$\sim$0.6\,\msun, \igmk\,$\sim$2.8, late-K spectral type). This hints at a possible relation between the dominant magnetic field topology and the convergence of stars onto a well-defined rotation sequence at a given age. It would be interesting to investigate this in more detail by comparing the modulation patterns and rotation distributions of clusters spanning a range of ages over the first billion years.

Our ability to detect evolution in the modulation patterns decreases with signal-to-noise (S/N) and is therefore harder for the fainter M stars. Specifically, it is harder to detect small levels of evolution as the relative noise level is higher for a given modulation amplitude. Significant evolution, which corresponds to the orange colours in Figure \ref{fig:P_col_lcm}, should be detectable for even the faintest stars in our sample. However, as we find that most M stars display stable modulation patterns, it is worth considering whether this finding is robust or is affected by our reduced ability to identify small evolution changes in low S/N light curves. To show this unambiguously would require injection-recovery tests in simulated \ngts\ light curves, which is beyond the scope of this work. We note, however, that similar conclusions have been postulated for M stars in the \pleiades\ \citep{Stauffer16}, based on \ktwo\ data with higher S/N for M0--M3 spectral types. We therefore suggest that the main conclusions on M stars here, i.e. that they often display stable modulation patterns over 200 day periods, are valid.

Finally, we note that \ktwo\ campaigns lasted 75 days and most \tess\ (Transiting Exoplanet Survey Satellite; \citealt{Ricker14}) fields will have 27 day coverage\footnote{With longer coverage towards the ecliptic poles where individual sectors overlap.}. Within a given 75-day \ktwo\ window, and even more so within a 27-day \tess\ window, it would be difficult to see evolution in the modulation shape to the extent that is evident in the \ngts\ data. Such a long temporal baseline, combined with this level of sampling and photometric precision, is unprecedented for young stars.

\subsubsection{Spot evolution and/or differential rotation?}

We noticed that for stars displaying significant evolution in their light curves, the strongest LS periodogram peak was often split into two close peaks or had a complex shape. Such split/complex LS peaks have been noted for similarly-aged young stars in the \pleiades\ based on analysis of \ktwo\ light curves \citep{Rebull16a}. This might tentatively suggest that two or more close periodic signals exist in the data, which could be a sign of differential rotation (under the assumption that spot modulation is sinusoidal and spot groups survive throughout most of our light curves) and/or evolution in the spot distributions. The GP MCMC posterior period distributions for such stars are well-defined and single-peaked, which would support the spot evolution scenario as the simplest explanation. However, we note that the fractional period uncertainties for these variable stars are typically larger than for stars displaying stable modulation patterns with similar periods and spectral types\footnote{We note, however, that given the small number of FGK stars displaying stable patterns, this assertion is based on low numbers.}. This is probably because the rotation periods of stars with variable modulation patterns are simply less well constrained (given the more complex nature of the variability and hence required flexibility of the GP model), but it could also be because the period distributions are actually the summation of two or more closely-overlapping, well-defined period distributions, as one might expect if differential rotation is present and significant enough to increase the spread in the measured rotation period distribution. Given that stars displaying evolving modulation patterns and broader rotation period distributions are typically FGK stars, this would require differential rotation to be more prominent in these stars than in M dwarfs. Indeed, measured rates of differential rotation have been observed to decrease with stellar temperature \citep[e.g.][]{Barnes05,CollierCameron07}, with measurements for M dwarfs at or below the fully convective boundary significantly lower than could be detected in the NGTS data of \blanco\ \citep[e.g.][]{Morin08a,Morin08,Reinhold13,Davenport15}. While we feel that the rotation periods of stars with evolving modulation patterns will simply be less well-constrained due to their more complex nature, it is certainly plausible that differential rotation is also present and stronger in the FGK stars.

Accurately decoupling spot evolution from differential rotation is troublesome (see e.g. \citealt{Aigrain15} for a discussion based on \kepler\ light curves of solar-type stars). Differential rotation has, however, been reported in young field stars observed by the \kepler\ prime mission \citep{Frasca11,Frohlich12}, based on multi-spot models that allow the shape of spots to evolve in time, but which assume the spots survive throughout the observations. The authors are not aware of modelling efforts that have been applied to young solar-type and low-mass stars that allow evolution in both the shape and existence of different active regions, such that a differential measure of the stellar rotation period is possible from each active region's modulation period during their lifetimes. Unfortunately, the precision and duration of the \ngts\ data, combined with our limited knowledge of spot lifetimes and behaviour at $\sim$100 Myr, do not allow us to distinguish between spot evolution and differential rotation with the current analysis.

\subsection{Colour--period--amplitude relation}
\label{sec:col_P_amp}

\begin{figure}
\centering
  \includegraphics[width=\linewidth]{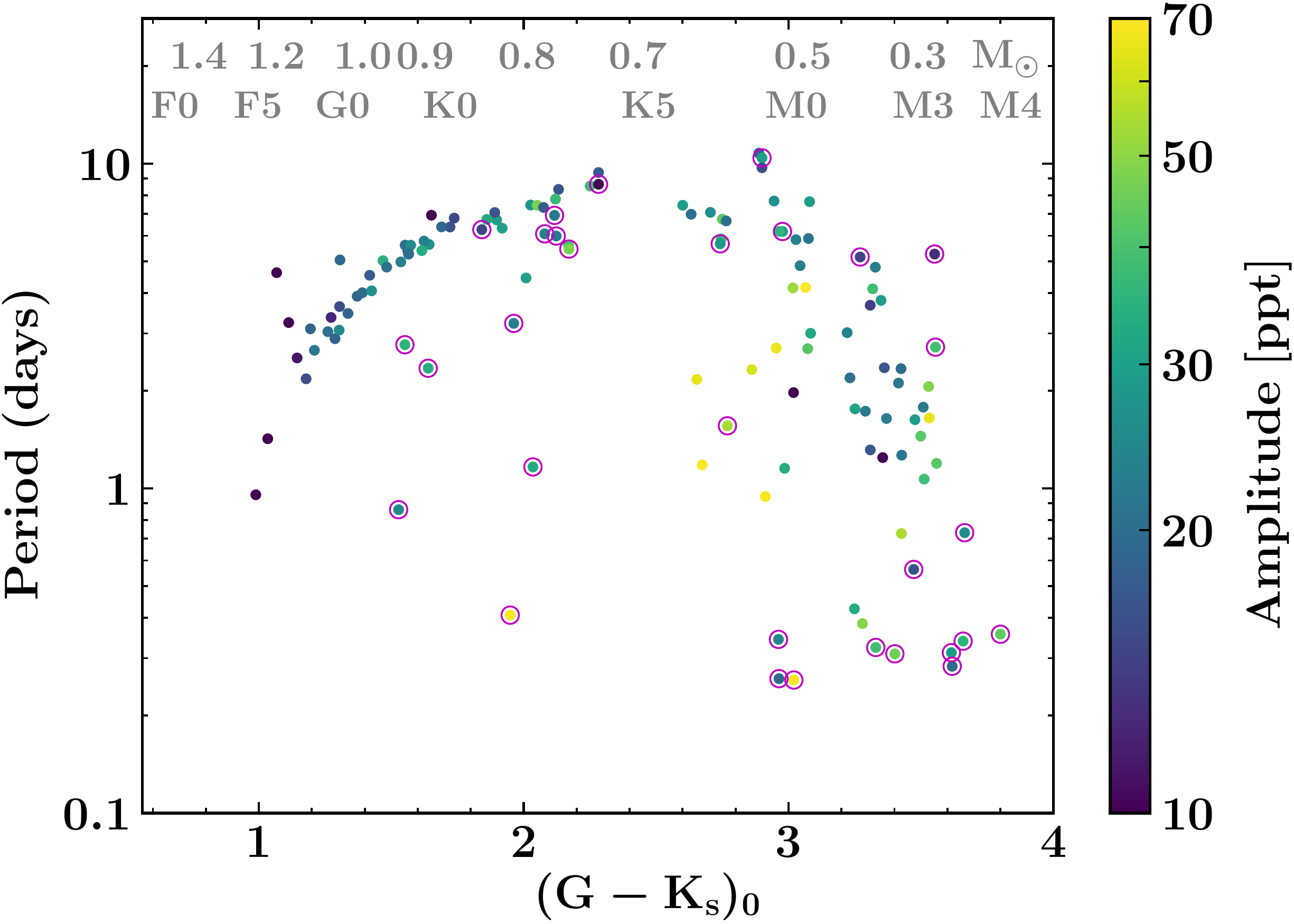}
    \caption{Rotation period vs. dereddened \gmk\ colour for stars in \blanco\ coloured by the amplitude of their light curve modulation patterns. Most notably for single late-K and early M stars (2.5\,$\lesssim$ \igmk\ $\lesssim$\,3.2), there appears to be a correlation between rotation period and modulation amplitude, with the faster rotators, which sit below the upper cluster envelope, displaying higher modulation amplitudes than their more slowly rotating counterparts. Photometric multiple stars are circled in magenta.}
    \label{fig:P_col_amp}
\end{figure}

Figure \ref{fig:P_col_amp} shows the colour--period relation for stars in \blanco\ coloured by the amplitude of their modulation patterns. Focusing on the single stars, as we move from mid-F to mid-K spectral types there is a slight increase in the average amplitude, which probably reflects the increasing size of the convective outer layer in these stars (see \S\ref{sec:col_amp} for further discussion). For the single late-K and early M stars (2.5\,$\lesssim$ \igmk\ $\lesssim$\,3.2), the faster rotators, which sit below the upper cluster envelope, display higher modulation amplitudes than their more slowly rotating counterparts. This is not exclusively the case, however, as some faster rotators display modest modulation amplitudes. If the inverse correlation between rotation period and modulation amplitude for single stars at a given colour (i.e. mass) is correct, these low-amplitude fast rotators may be either: 1. stars that were observed at comparatively low inclination angles; or 2. binary systems that have lower mass-ratios than we could detect with our CMD analysis. We note that the binaries we did identify do not show an obvious trend between rotation period and amplitude at a given \igmk\ colour. This is unsurprising as each system will have different modulation signals from the component stars, which may differ in amplitude (depending on the light ratio between the components), and constructively and/or destructively interfere over the course of the \ngts\ observations (if the components are not tidally locked with negligible differential rotation). 

We refrain from postulating about the lowest mass stars (\igmk\,$\gtrsim$\,3.2) because we become increasingly more sensitive to higher amplitude variables for the faintest stars.


\section{Comparing Blanco 1 and the Pleiades}
\label{sec:B1_plei_comp}

\begin{figure*}
\centering
  \includegraphics[width=0.48\linewidth]{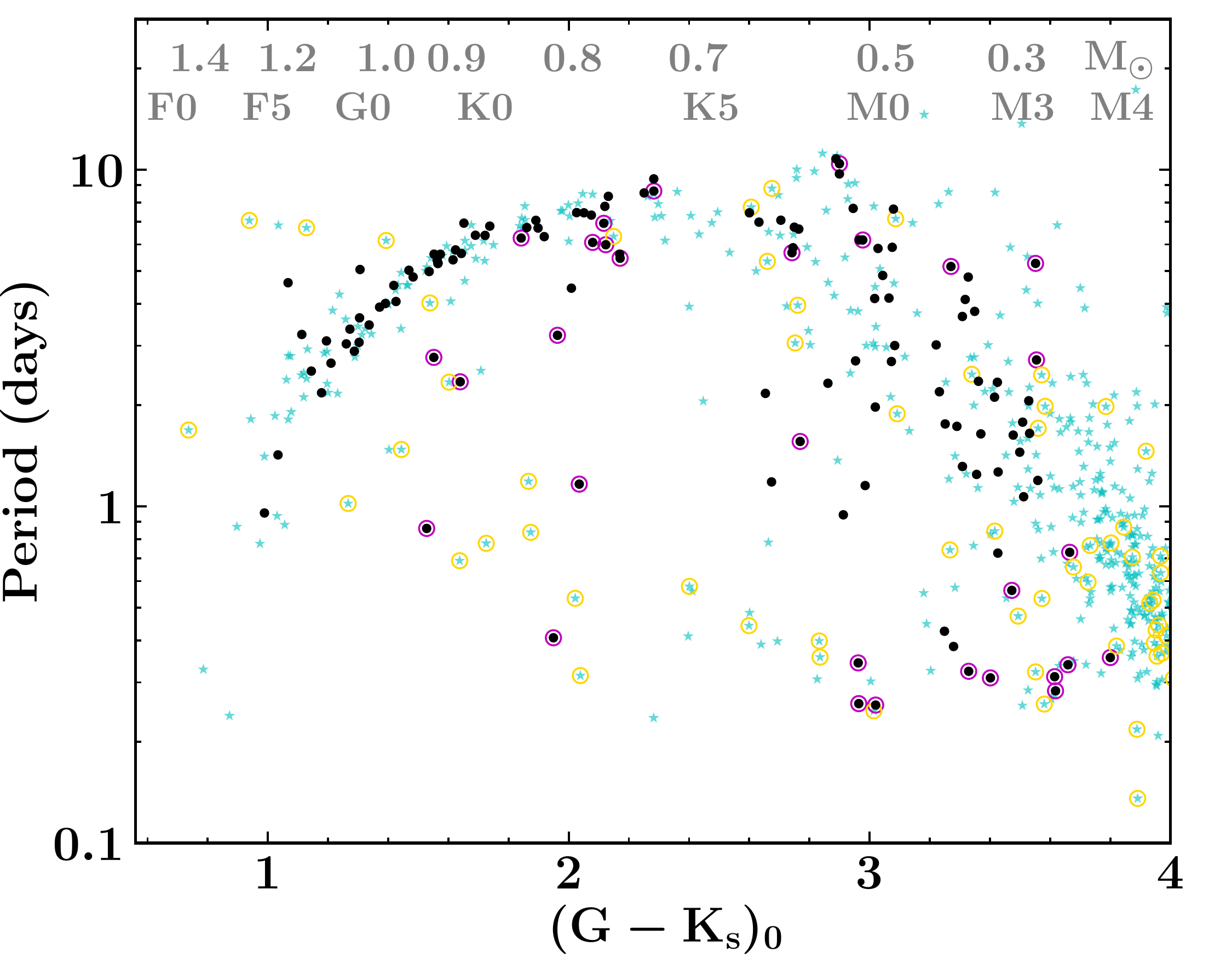}
  \includegraphics[width=0.48\linewidth]{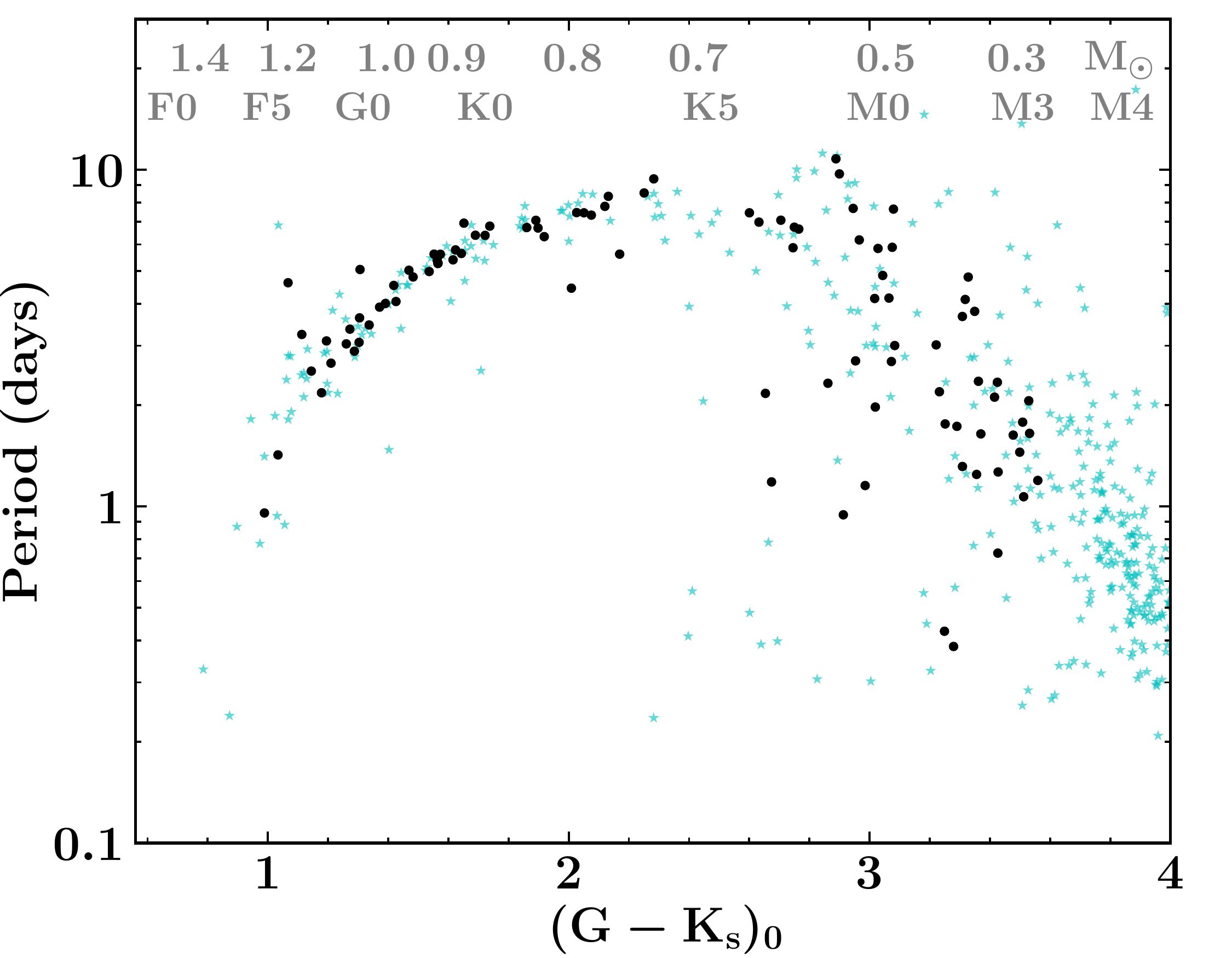}
    \caption{Rotation period vs. dereddened \gmk\ colour for stars in \blanco\ and the \pleiades. This is the same as Figure \ref{fig:P_col} with the addition of the \pleiades\ data, which has been dereddened assuming E(B-V) = 0.045 for the cluster (\colblue{B18}). \emph{Left}: all stars with detected rotation periods (black points for \blanco\ and cyan stars for the \pleiades), with our identified multiple star systems highlighted (magenta circles for \blanco\ and gold circles for the \pleiades). \emph{Right}: showing only the apparently single stars to highlight the clear rotation sequences between 1.1$<$\igmk$<$2.3 (1.2$\,\gtrsim$\,$M$\,$\gtrsim$\,0.75\,\msun). Mass-dependent angular momentum evolution is strongly imprinted in both clusters.}
    \label{fig:P_col_B1P}
\end{figure*}

\subsection{Colour--period distribution}

Stellar rotation in the \pleiades\ has been extensively studied \citep[e.g.][]{Hartman10,Covey16,Rebull16,Rebull16a,Stauffer16} and now \blanco\ also possesses a well-constrained distribution of rotation periods. Given the very similar age estimates for these clusters, we compare their rotation period distributions to understand the level to which mass-dependent angular momentum loss mechanisms are imprinted on stars at $\sim$110--115 Myr.

We cross-matched the \gaia-DR2 \pleiades\ membership list with stars from \citet{Rebull16}, which resulted in 589 Pleiads with measured rotation periods\footnote{The \citet{Rebull16} sample contained 759 likely members. While some of the stars not in the  \gaia-DR2 membership list may be non-members, many are likely Pleiads that reside in multiple star systems, which may be affecting the DR2 astrometric solution. It would be interesting to revisit these stars in future data releases.}. 
We also performed our CMD fitting technique on the \pleiades\ to identify likely multiple systems. The colour-period distributions of both clusters are shown in Figure \ref{fig:P_col_B1P} (full distributions in the left hand panel and only the apparently single stars in the right hand panel).

The period distributions in both clusters are strikingly similar. Figure \ref{fig:P_col_B1P} (right hand plot) shows that essentially all single stars in the mid-F to mid-K spectral range (0.7\,$\lesssim$\,$M$\,$\lesssim$\,1.2 \msun) lie on tight cluster sequences. This rotation sequence is slightly tighter for \blanco\ than for the \pleiades, which we attribute to a combination of a longer observation baseline (200 vs. 75 days) and the novel period estimation methods applied here.

\subsubsection{FGK stars with intermediate rotation periods}

In addition to the tight cluster sequence, previous \pleiades\ rotation studies identified two populations of stars below this sequence, termed intermediate and fast rotators based on the magnitude of their displacement below the main cluster sequence (see e.g. Figure 2 in \citealt{Stauffer16}). Following our \gaia\ DR2-based membership selection from \colblue{B18}, we do not see the intermediate sequence as strongly, which suggests that either these were mainly non-members or their \gaia\ DR2 astrometric solutions were suspect, which could be due to the effect of binarity, as \citet{Stauffer16} noted these stars were preferentially displaced above the CMD cluster sequence.

There are a few stars that remain in this intermediate rotator sequence, however, which can be seen sitting just below the well-defined single-star sequence (2 in \blanco\ and 2 in the \pleiades). We suspect these are likely binaries with low-mass companions and therefore were not flagged by our photometric identification methods. It is interesting to note that these stars appear to follow a trend with rotation periods $\sim$2--3 days shorter than the main rotation sequence, so intermediate `sequence' is perhaps a relevant term. However, with only four such stars, we refrain from making further statements; additional clusters with similar ages are needed to shed further light on this potential small sub-population.

\subsubsection{The apparent kink in the \pleiades\ single star sequence around a spectral type of $\sim$K5}

The \pleiades\ data in Figure \ref{fig:P_col_B1P} shows an apparent kink in the single star sequence around a spectral type of K5 (\igmk\,$\sim$2.5), with the upper envelope of the late K stars reaching longer rotation periods than earlier spectral types. This kink was first noted by \citet{Stauffer16}. Here, we see the kink even more clearly following the \gaia\ DR2 membership selection, suggesting that it is probably a real phenomenon within the cluster sample (assuming the rotation periods for these stars are accurate). There is a dearth of late K stars in \blanco, which means we cannot strongly comment on the presence of such a kink in this cluster, although we note that the \blanco\ rotation period distribution within this mid-to-late K spectral range (2.3\,$\lesssim$\,\igmk\,$\lesssim$\,2.9) is consistent with the \pleiades, and hence also consistent with the presence of such a kink.

\subsection{Fitting the single FGK star cluster sequences}

\begin{figure}
	\includegraphics[width=\columnwidth]{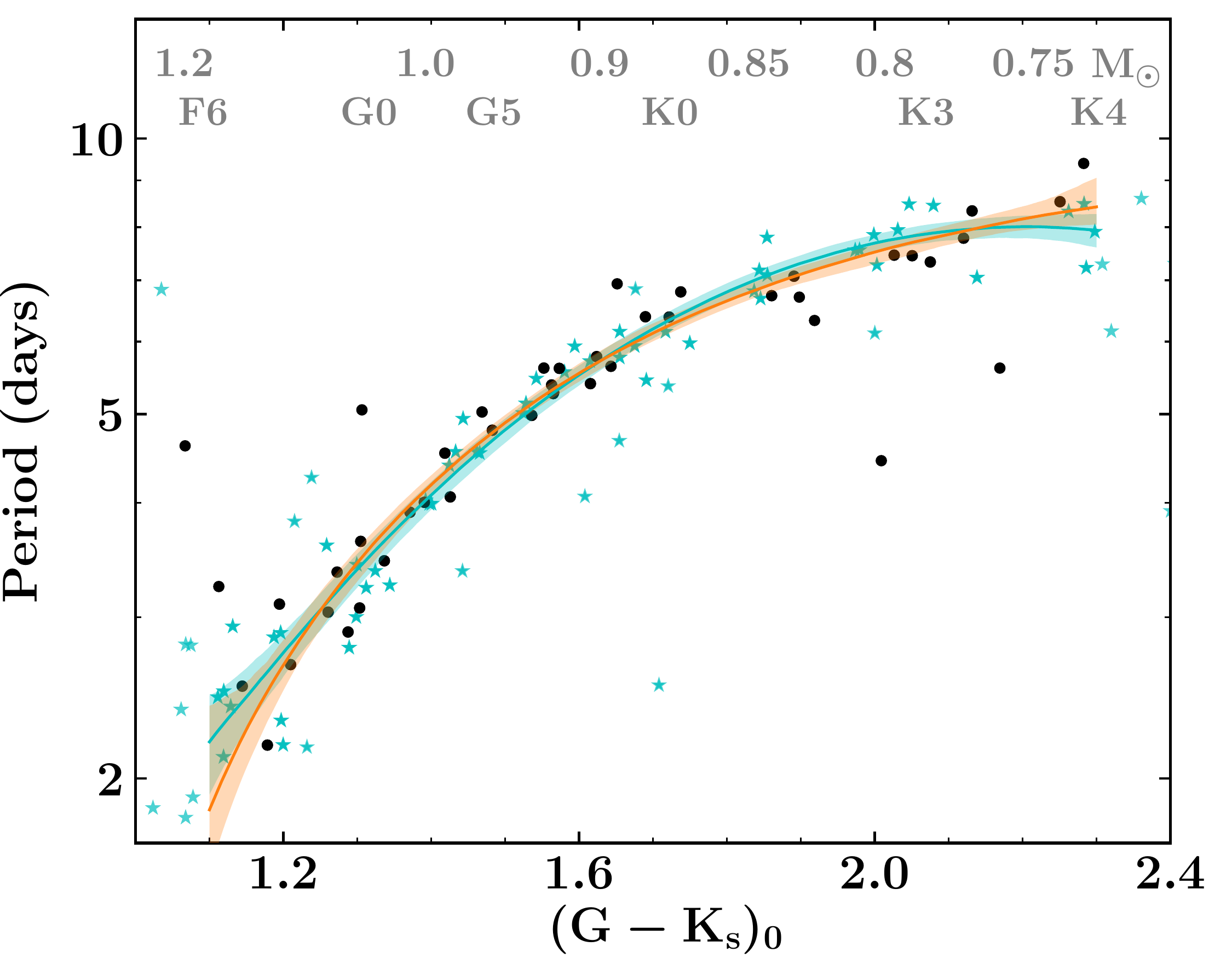}
    \caption{Gaussian process (GP) fits to the single star rotation sequences in \blanco\ (black points, orange model) and the Pleiades (cyan stars, cyan model). The lines and shaded regions indicate the mean and 2$\sigma$ confidence interval of the posterior GP distribution. The rotation sequences of \blanco\ and the \pleiades\ are consistent to within their uncertainties across the mid-F to mid-K spectral range (1.2$\,\gtrsim$\,$M$\,$\gtrsim$\,0.75\,\msun).}
    \label{fig:SS_rot_fit}
\end{figure}

We fit the rotation periods of single mid-F to mid-K stars in \blanco\ and the \pleiades\ to assess the similarity between the rotation sequences in these two clusters. For this, we focus on stars between 1.1\,$<$\,\igmk\,$<$\,2.3, as these ranges encompass the most well-defined section of the rotation sequences, and are well-populated by stars in both clusters.

Figure \ref{fig:SS_rot_fit} shows our fits to both cluster sequences. We opted to use a GP model because this marginalises over an ensemble of functions, which is more general than using gyrochronology or polynomial models, but also includes (models similar to) them. We use a Squared Exponential kernel, as implemented in {\tt george} \citep{Ambikasaran15} and perform a two-stage fit, first running an iterative maximum a posteriori fit with 3$\sigma$ outlier rejection, before running an MCMC fit to each cluster sequence (5000 steps, 200 walkers, with a 3000 step burn-in) using {\tt emcee} \citep{Foreman-Mackey13}. Iterative outlier rejection was performed, even though we fit only the (apparently) single star sequences, because our multiple star identification methods are not sensitive to all multiple star mass ratios and separations. It is possible, therefore, that some of the fitted stars reside in multiple star systems where their angular momentum evolution has been affected by hitherto unidentified companions, and hence should not be included in this analysis, which is seeking to focus on the single star rotation sequences in both clusters.

It is clear that the \blanco\ and \pleiades\ rotation sequence fits (orange and cyan lines and shaded regions) are consistent with each other\footnote{Performing a similar fit with quadratic polynomial models also suggests the two rotation sequences are consistent, although a less extensive range of models are explored.}. It follows that the angular momentum evolution of mid-F to mid-K stars follows a well-defined pathway that is strongly imprinted by $\sim$100\, Myr, irrespective of their individual angular momentum evolution histories. Furthermore, CMD isochronal age estimates for both clusters are essentially identical (to within $\sim$5\,Myr in \colblue{B18}). From their rotation sequence agreement, it follows that their gyrochronological ages would also agree, which is quite encouraging for gyrochronolgy relations seeking to identify a singular rotation-mass-age relationship that has thus far proved elusive \citep{Angus15}. We note, however, that the scatter in the two sequences is larger than the posterior confidence intervals, which suggests that there is some intrinsic scatter in the data above a singular period-colour relation. Performing a detailed gyrochronological modelling of the \blanco\ and \pleiades\ single star sequences is beyond the scope of the present paper; we leave this to future work.

\subsection{Assessing the similarity between the low-mass populations}

While it is intuitively straightforward to show that the mid-F to mid-K stars in \blanco\ and the \pleiades\ follow a consistent trend (to within the precision of the current rotation period data), it is harder to show this for the lower mass populations. This is primarily because: 1. the low-mass stars do not follow a well-defined sequence; 2. multiple star contamination will likely be higher because our identification methods are less sensitive for M stars, as they possess an intrinsically broader spread in luminosities; and 3. the \ngts\ observations of \blanco\ do not probe as deep as the \ktwo\ \pleiades\ data, which means we are progressively more sensitive to larger amplitude variables in \blanco\ compared to the \pleiades\ for the latest spectral types. Given these complicating factors, we opt to carry out a simpler test of similarity between the two populations.

We perform 2-sample KS (Kolmogorov-Smirnov) and AD (Anderson-Darling) tests \citep[e.g.][]{Feigelson12} on the period distributions of the low-mass single stars (2.6\,$<$\,\igmk\,$\leqslant$\,3.6). Both KS and AD tests are non-parametric and distribution free. The main difference, in the context of the present work, is that AD is more sensitive to differences near the edges of distributions.

We note that neither KS or AD tests are strictly valid when data are distributed in two (or more) dimensions, as is the case here, because there is no unique way to order the data. This means that two datasets can yield the same empirical distribution function (EDF) while possessing different distributions, thereby invalidating the KS or AD test statistics. To account for this, we perform KS and AD tests on data within small $\Delta$\,\igmk\,$=$\,0.2 colour slices, on the basis that any trend within such a small colour range is negligible given the spread and sampling of the data. Under this assumption, we find that both the KS and AD tests cannot reject the null hypothesis that the \blanco\ and \pleiades\ low-mass rotation period distributions are drawn from the same parent population. The AD test is generally less confident than the KS test, given that it is more sensitive to stars with short and long periods, and there appears to be a dearth of fast ($\sim$0.3--0.6 day) single star rotators in \blanco\ compared to the \pleiades\ in this mass range.

\subsection{Probing starspot distributions across FGKM stars}
\label{sec:col_amp}

\begin{figure}
\centering
  \includegraphics[width=\linewidth]{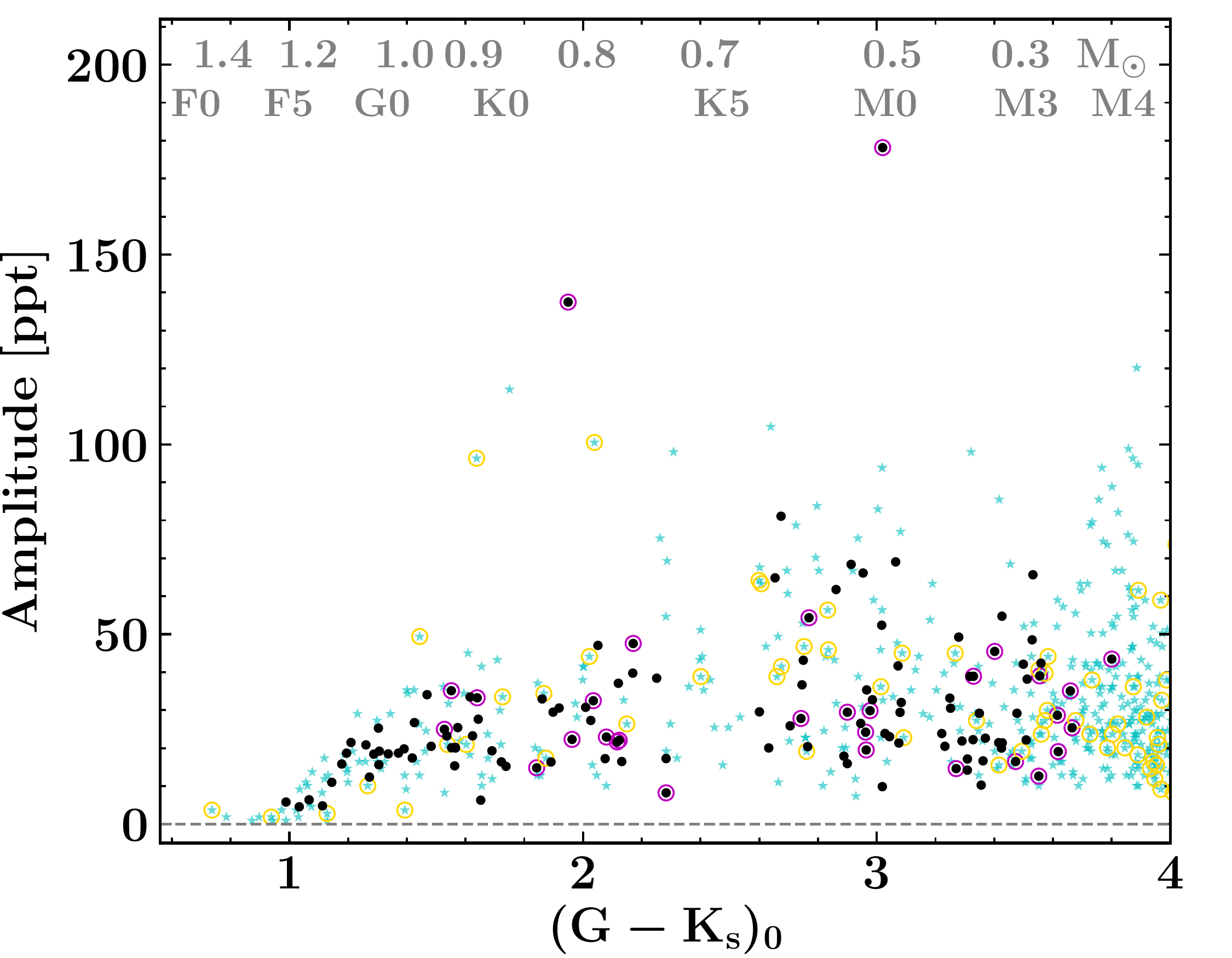}
    \caption{Colour-amplitude relation for \blanco\ and the \pleiades. \blanco\ stars are represented by black points, with multiple stars circled in magenta, and Pleiads are indicated by cyan stars, with multiples circled in gold. There is no clear distinction between the modulation amplitudes of single and multiple stars, except that very large amplitude variables are preferentially found in multiple systems. Stellar masses (\msun) and spectral types are indicated at the top. At a spectral type of $\sim$F5, variability amplitudes start to increase, which we attribute to the emergence of sufficiently deep convective envelopes that can drive and sustain a significant magnetic dynamo, and hence give rise to the starspot distributions whose longitudinal inhomogeneity drives the observed modulation patterns.}
    \label{fig:amp_col_B1P}
\end{figure}

Figure \ref{fig:amp_col_B1P} shows the measured amplitudes of the modulation patterns as a function of \igmk\ colour for both \blanco\ and the \pleiades. Amplitude is defined here as the 10$^{\,\rm th}$--90$^{\,\rm th}$ flux percentile of the GP model (\blanco) and the data (\pleiades)\footnote{We opted to use our GP model prediction rather than the \blanco\ data itself because the \ngts\ data is not continuous whereas the GP model is, which makes it a more appropriate comparison to the continuous space-based \pleiades\ data.}. Moving from left to right on the plot, the amplitude of photometric variability shows a clear increase around a spectral type of F5 (\igmk$\sim$1.0). We attribute this to the emergence of sufficiently deep convective envelopes \citep{Wilson66}, which drive the magnetic dynamos that in turn give rise to surface starspot distributions (whose longitudinal inhomogeneity drives the observed rotational modulation patterns). 
This increase in modulation amplitude around F5 spectral type was also noted in the \pleiades\ \citep{Rebull16}.
Once convective envelopes become sufficiently deep (\igmk\,$\sim$\,1.4, early G spectral type), there is no clear trend between modulation amplitude and stellar mass. The amplitudes of mid-G to mid-M stars are predominantly spread between 10--50 ppt with a scattering of higher amplitude variables. 
However, as shown in Figure \ref{fig:P_col_amp}, late-K and early M stars (2.5\,$\lesssim$ \igmk\ $\lesssim$\,3.2) in \blanco\ show a tentative trend where faster rotators, which sit below the upper cluster envelope, typically display higher modulation amplitudes than their more slowly rotating counterparts. This is the primary driver for the scattering of higher amplitude variables in this colour range in Figure \ref{fig:amp_col_B1P}.

Finally, we note that the variability observed in the four earliest spectral type stars in \blanco\ (\igmk\,$\lesssim$\,1.1) could result from pulsations rather than rotational modulation. ZAMS F stars have temperatures that allow a range of pulsational variability (e.g. $\delta$ Scuti and $\gamma$ Doradus), with $\gamma$ Dor stars possessing pulsation periods (0.4--3.0 days) that overlap with the single star rotation sequence at this mass \citep{Kaye99,Balona11,Stauffer16}. Given their low variability amplitudes, we cannot easily distinguish between the two variability mechanisms for these stars.


\section{Conclusions}
\label{sec:conclusions}

We conducted a $\sim$200 day photometric monitoring campaign of the $\sim$115 Myr old \blanco\ open cluster with NGTS. We determined rotation periods for \nProt\ stars spanning F5--M3 spectral types (0.3\,$\lesssim$\,M\,$\lesssim$\,1.2\,\msun), which increases the number of rotation periods in the cluster by a factor of 4. 

We used three independent methods to estimate rotation periods: Gaussian process (GP) regression, generalised autocorrelation (\gacf) and Lomb-Scargle (LS) periodograms. We find that the GP and \gacf\ methods are better suited to estimating rotation periods for solar-type stars that display evolving modulation patterns, as these methods are more flexible than LS. All three methods perform well for stars with stable modulation patterns. In addition to estimating rotation periods, we identified binary and higher order multiple star systems by fitting the cluster sequence in colour-magnitude space and cross-matching with literature RV surveys.

The rotation period distribution of F5--M3 stars in \blanco\ shows three main features: 1. single stars between mid-F and mid-K follow a well-defined rotation sequence from $\sim$2 to 10 days; 2. the photometric multiples within this spectral type range typically sit below the single star sequence with shorter rotation periods; and 3. the late-K and M stars possess a broader spread of rotation periods between $\sim$0.3--10 days with multiple stars spread throughout this distribution.

The fact that mid-F to mid-K photometric multiples have faster rotation rates than their single star counterparts may suggest that the presence of a close companion with a moderate-to-high mass ratio inhibits angular momentum loss mechanisms during the early pre-main sequence, and this signature has not been erased at $\sim$100 Myr. 

We find that the majority of mid-F to mid-K stars display modulation patterns that show moderate-to-significant evolution in amplitude and/or phase shape. In contrast, most M0--M3 stars appear to possess reasonably stable modulation signals. This difference could arise from different dominant magnetic field morphologies in mid-F to mid-K stars compared to M dwarfs at this age, with the transition occurring at late-K spectral types. Interestingly, this morphological change coincides with the shift from a well-defined rotation sequence (mid-F to mid-K stars) to a broad rotation period distribution (M stars) at this age. This hints at a possible relation between magnetic field topology and convergence onto a well-defined rotation sequence at a given age.

Finally, we compared our rotation period distribution for \blanco\ to the similarly-aged \pleiades. We find that the single star populations in both clusters possess consistent rotation period distributions, which suggests that the angular momentum evolution of stars follows a well-defined pathway that is, at least for mid-F to mid-K stars, strongly imprinted by $\sim$100 Myr. This is quite encouraging for gyrochronolgy relations seeking to identify  a  singular  rotation-mass-age relationship.

\section*{Acknowledgements}
EG thanks Cristina Blanco for insight that helped shape the direction of this paper, and Richard Booth, Mihkel Kama, Laetitia Delrez, Suzanne Aigrain, John Stauffer and Luisa Rebull for enjoyable and informative discussions.
This research is based on data collected under the NGTS project at the ESO La Silla Paranal Observatory. The NGTS facility is funded by a consortium of institutes consisting of 
the University of Warwick,
the University of Leicester,
Queen's University Belfast,
the University of Geneva,
the Deutsches Zentrum f\" ur Luft- und Raumfahrt e.V. (DLR; under the `Gro\ss investition GI-NGTS'),
the University of Cambridge, together with the UK Science and Technology Facilities Council (STFC; project reference ST/M001962/1).
EG gratefully acknowledges support from the David and Claudia Harding Foundation in the form of a Winton Exoplanet Fellowship.
PJW is supported by STFC consolidated grant ST/P000495/1.
CAW acknowledges support by the STFC grant ST/P000312/1.
Finally, we would like to thank the anonymous referee for their insightful reading of the manuscript and helpful suggestions for improvement.




\bibliographystyle{mnras}
\bibliography{ref}


\appendix
\section{Removing residual moon variations arising from incomplete background correction}
\label{sec:appendix}

During the course of the 200 day \ngts\ observations, the moon passes through several lunar cycles, with corresponding brightness variations. These are corrected for within the standard \ngts\ pipeline through a \textsc{sysrem}-based detrending algorithm \citep[see][for more details]{Wheatley18}. This works well for all but the faintest stars (\ngts\ mag $\gtrsim$15 mag), which typically still display a level of variability in phase with the lunar cycle. We define moon signal here as periodic variability between 25$<$P$<$30 days, whose phase shape coincided with the moon's brightness variations during each lunar cycle. In some cases, these variations have an amplitude comparable to the rotation signals we are trying to detect. We opted, therefore, to remove the residual moon signal, where applicable, during our procedure for estimating rotation periods. The procedure is described below.

For each light curve, we performed an initial Lomb-Scargle (LS) fit. If moon signal was detected in the four strongest non-aliased peaks, it was removed. The removal process comprised two steps. The light curve was folded on the detected `moon' period and detrended for the dominant variation pattern using a Savitzky-Golay (SG) filter followed by a convolution. We opted to perform this two-step process as it allowed us to best capture the shape of the moon signal on the first pass (SG filter) and then smooth this signal (via convolution) to apply a smooth detrending of the moon signal.

In practice, the residual moon signal was only apparent in stars fainter than $\sim$14.5 mag in the \ngts\ band, which corresponds to early M stars and later spectral types in \blanco. We checked that removing the moon signal did not significantly change the rotation periods determined for a handful of example light curves, with signals across a range of amplitudes. For some low-mass stars, however, it allowed us to determine rotation periods with a higher degree of confidence.

\bsp	
\label{lastpage}
\end{document}